\documentclass[%
 notitlepage, 10pt, twocolumn,
 amsmath,amssymb,
 aps,prx,
floatfix,
]{revtex4-2}

\usepackage{graphicx}
\usepackage{dcolumn}
\usepackage[colorlinks, citecolor=blue, linkcolor=red]{hyperref}
\usepackage{bm}
\usepackage{color} 
\usepackage[caption = false]{subfig}
\newcommand{\ket}[1]{|#1\rangle}

\begin{document}

\preprint{APS/123-QED}

\title{Optimal time for sensing in open quantum systems}

\author{Zain H. Saleem}
\email{zsaleem@anl.gov}

\affiliation{Mathematics and Computer Science Division, 
Argonne National Laboratory, 9700 S Cass Ave, Lemont IL 60439}
\author{Anil Shaji}
\email{shaji@iisertvm.ac.in}

\affiliation{School of Physics, IISER Thiruvananthapuram, Kerala, India 695551}

\author{Stephen K. Gray}
\email{gray@anl.gov}

\affiliation{Center for Nanoscale Materials,
Argonne National Laboratory, Lemont, Illinois 60439, USA}

\date{\today}

\begin{abstract}
We study the time-dependent quantum Fisher information (QFI) in an open quantum system satisfying the Gorini-Kossakowski-Sudarshan-Lindblad master equation. 
We also study the dynamics of the system from an effective non-Hermitian dynamics standpoint and use it to understand the scaling of the QFI when multiple probes are used.
A focus of our work is how the QFI is
maximized at certain times suggesting that the best precision
in parameter estimation can be achieved by focusing on these times. The propagation of errors analysis allows us to confirm and better 
understand this idea.
We also propose a parameter estimation procedure involving relatively low resource consuming measurements followed by higher resource consuming measurements and demonstrate it in simulation.
\end{abstract}

\maketitle

\section{\label{sec:level1} Introduction}

Quantum sensing and metrology \cite{braunstein1994statistical,giovannetti_quantum-enhanced_2004,degen2017quantum,braun_quantum-enhanced_2018,pezze_quantum_2018,giovannetti2006quantum,giovannetti_advances_2011,barbieri_optical_2022} involve the exploration of subtle quantum effects to increase the precision of parameter estimation. 
Quantum sensing has become one of the most promising applications of quantum technologies, involving single- or multi-parameter estimation. In this work we will use quantum Fisher information (QFI) as a tool to study quantum sensing for open quantum systems \cite{faist2021time,alipour2014quantum,naghiloo2019quantum,lu2010quantum,gammelmark2014fisher,altintas2016quantum}. The QFI \cite{braunstein1994statistical} quantifies the theoretical bound on the achievable precision in estimating a parameter using a quantum state as a probe and can be regarded as a performance measure of a quantum system as a quantum sensor. The open quantum systems we will study in this work are dynamic, i.e., evolve with time, and therefore it makes more sense to study the time dependence of the QFI. 

We consider as quantum probes one or more two-level systems or qubits and employ two different approaches to study their environmental interactions or open system dynamics  ~\cite{braunstein1994statistical}. The first is based on the Gorini-Kosskowski-Sudarshan-Lindblad (GKSL) master equation~\cite{gorini_completely_1976,lindblad_generators_1976,chruscinski_brief_2017,manzano2020short} where we assume a Markovian interaction of the probe with its environment and integrate out the degrees of freedom of the environment to derive a dynamical equation for the probe. The second approach is based on a non-Hermitian extension of quantum mechanics  \cite{ashida2020non} and allows us to investigate sensors with a large number of probes. Here the Hamiltonian describing the evolution  of the probe is assumed to acquire an anti-Hermitian part which can be associated with dissipative effects. 
One of our results is to show how the two-level non-Hermitian systems can be extended to display the GKSL dynamics. 

In the non-Hermitian approach one also can encounter exceptional points that mark the transition of the Hamiltonian from a PT-symmetric form to one that is not PT-symmetric \cite{bender1999pt}. Possible quantum advantages in sensing and metrology facilitated by such exceptional points have been of interest recently \cite{lau2018fundamental}. Framing the metrology scheme using the non-Hermitian as well as GKSL master equation based approaches also allow us to address the question of metrological advantage around the exceptional points. We find no such advantage at the exceptional point, consistent
with several previous studies of related systems \cite{Langbein2018,lau2018fundamental}.

In the absence of dissipation, the QFI for the system we consider increases monotonically with time. 
This means that the achievable precision in the estimate of the parameter of interest will improve with increased duration of the measurement. However, when dissipation is present, this is not the case. We find that for the parameter estimation problem we are considering, there is an optimal time at which QFI is largest and consequently one can expect to get best possible measurement precision at this time. With dissipation, it is important to also verify whether the bound on the measurement precision given by the QFI is achievable in practice. 

In section \ref{one} we will introduce classical and quantum Fisher information. We investigate the time dependent QFI for the open quantum systems via the GKSL formalism in section \ref{three} and discuss the extension to N-probes case via the non-Hermitian approach.  In section \ref{expt} we compare our result obtained via the time dependent QFI with the propagation of error in the variance of the parameter and show that they both match to a high accuracy. We suggest an experimental procedure for parameter estimation making use of the optimum time concept in section \ref{optimal}. Finally in section \ref{six} we give conclusions and future directions.

\section{Quantum Fisher Information and Parameter Estimation}\label{one}
Let us consider the problem of simultaneously estimating $n$ parameters $\textbf{x}_i=\{x_1,x_2, \cdots x_n\}$ in a quantum experiment and denote the respective estimators of these parameters by $\hat{\textbf{x}}_i=\{\hat{x}_1,\hat{x}_2, \cdots \hat{x}_n\}$. The uncertainty in the the estimator $\hat{\textbf{x}}_i$ is quantified by the covariance matrix $\text{Cov} (\hat{\textbf{x}}_i, \hat{\textbf{x}}_j)$ and is upper bounded by the quantum Cramér-Rao Bound~\cite{holevo2011probabilistic}, 
\begin{equation}
  \text{Cov} (\hat{\textbf{x}}_i, \hat{\textbf{x}}_j) \geq \frac{1}{M F_{i,j}}  
\end{equation}
where $M$ stands for the total number of experiments and $F_{i,j}$ is the quantum Fisher information matrix which quantifies the responsiveness of the quantum state of the probe to changes in the measured parameters $\textbf{x}_i$. The coefficients of the QFI matrix for a given initial state of the probe given by the density matrix $\rho$ are given by the formulae,
\begin{equation}
    \label{fisher}
    F_{i,j}= \text{Tr}\bigg({L_i \frac{\partial{\rho}}{\partial x_j}}\bigg),
\end{equation}
where $L_i$ is the symmetric logarithmic derivative~\cite{helstrom1969quantum} for the parameter $x_i$ and is defined implicitly by,
\begin{equation}
    \frac{\partial{\rho}}{\partial x_i} =\frac{1}{2}( \rho L_i + L_i \rho )~~.
\end{equation}
Upon writing the density matrix in its eigenbasis  $\rho = \sum_a \lambda_a |\lambda_a\rangle \langle \lambda_a |$ and substituting in the above equation we get, 
\begin{equation}
    L_{i} = \sum_{ \{  a b| \lambda_a+\lambda_b \neq 0 \}} \frac{2}{\lambda_{a}+\lambda_{b}} \langle\lambda_{a}|\partial_{i}\rho |\lambda_{b}\rangle |\lambda_a\rangle \langle \lambda_b|,
\end{equation}
This formula for the SLD when substituted in \eqref{fisher} gives us the QFI for a mixed state \cite{liu2019quantum},
\begin{equation}
    \label{fisher1}
F_{ij}=\sum_{ \{  a b| \lambda_a+\lambda_b \neq 0 \} } F_{ij}(a,b)
\end{equation}
where,
\begin{equation}
\label{indicomp}
F_{ij}(a,b)=
\frac{2\mathrm{Re} \big[ \langle\lambda_{a}|\partial_{i}\rho |\lambda_{b}\rangle\langle\lambda_{b}|\partial_{j}\rho |\lambda_{a}\rangle \big]}{\lambda_{a}+\lambda_{b}},
\end{equation}
In the above formula one needs to ensure that the summation is performed only over those eigenvalues for which ${\lambda_{a}+\lambda_{b}} \neq 0$. With a little bit of work one can also show that if we have pure parameterized quantum state $|\psi\rangle:=|\psi(\vec{x})\rangle$
then the formula for the QFI matrix becomes,
\begin{equation}\label{pure}
F_{ij}=4\mathrm{Re} \big[ \langle\partial_{i}\psi|\partial_{j}\psi\rangle-
\langle\partial_{i}\psi|\psi\rangle\langle\psi|\partial_{j}\psi\rangle \big].
\end{equation}
The Fisher information for a particular parameter $x_i$ is defined as $F \equiv F_{i,i}$. Since we will be exclusively considering single parameter estimation we use $F$ for QFI and $F(a,b)$ for the individual elements in the mixed state QFI formula, Eq.~\eqref{indicomp}.

\section{Time Dependent Quantum Fisher Information }\label{three}

\subsection{Single Probe Case}
Let us consider a three-level system with a Hamiltonian in the interaction
picture or rotating frame given by,
\begin{equation}
    H = \hbar g ( |e\rangle \langle f| + |f\rangle \langle e|) + \hbar \Delta (|f\rangle \langle f| - |e\rangle \langle e|)
     \label{hamil10}
\end{equation}
where $|e \rangle$  and $|f \rangle$ are the first two excited states of the system, $g$ is the coupling between these excited states and $\Delta$ is the detuning parameter. The third state which is the `sink' $|s\rangle$ enters the dynamics through the  GKSL master equation,
\begin{equation}
\dot{\rho} = \frac{1}{i\hbar} [ H, \rho ] +  L_e\rho L_e^{\dagger} - \frac{1}{2} \{L_e^{\dagger}L_e , \rho \}.
\label{Lindblad2}
\end{equation}
The jump operator describing the dissipation is 
\[ L_e=\sqrt{\gamma_e} |s\rangle \langle e| \]
and it is responsible for the loss of energy from state $|e\rangle$ to the sink, $|s\rangle$. We note that Eqs.~(\ref{hamil10}) and (\ref{Lindblad2}) and non-Hermitian variations have
previously been considered by Murch and co-workers \cite{naghiloo2019quantum, chen2021}
as models of a three-level superconducting transmon qubit. Of course, this simple
model could also describe many other situations, for example, the single excitation
manifold of a two-level qubit coupled to a bosonic cavity in a Jaynes-Cummings model.

\begin{figure*}[!htb]
    \subfloat{\includegraphics[width=0.5\textwidth]{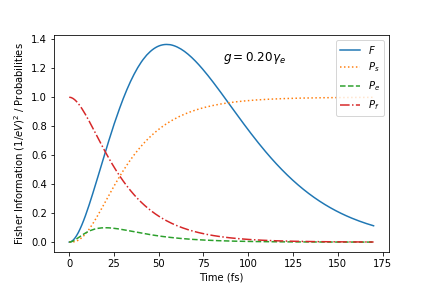}} 
    \subfloat{\includegraphics[width= 0.5\textwidth]{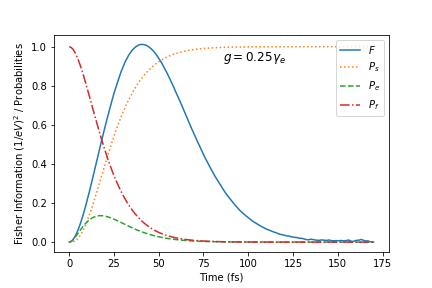}} \\[-5 mm]
    \subfloat{\includegraphics[width= 0.5\textwidth]{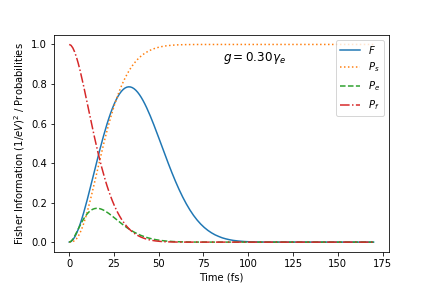}} 
    \subfloat{\includegraphics[width= 0.5\textwidth]{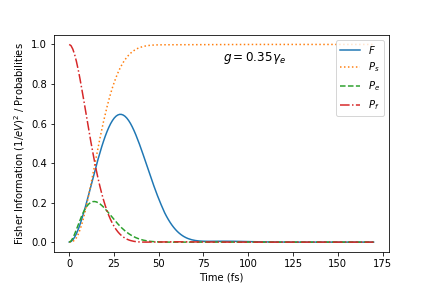}} 
    \caption{ The Quantum Fisher Information $F$  corresponding to estimation of the coupling $g$ are plotted as functions of time for different values of $g$. The initial state of the quantum probe is $|f\rangle$ and the collapse operator used is $L_e$. The values of $g$ are shown in the respective figures while the other parameters used are: $ \hbar \gamma_e = 0.150$ eV and $\hbar \Delta=0.0$ eV. Also shown in the figures are the density elements $\rho_{\rm ee}$, $\rho_{\rm ff}$ and $\rho_{\rm ss}$ of the quantum probe. \label{Lindbladf}} 
\end{figure*}

It is instructive to connect this model to the physical scenario of estimating the interaction between an atom and a cavity mentioned earlier.~Imagine a stream of atoms passing through the cavity one by one. Assume that initially the atom is in a state $|f\rangle$ and that the cavity induces coherent transitions between $|f\rangle$ and another atomic level $|e\rangle$.  In the lossless case, the probability of the atom exchanging a photon with the cavity and coming to the state $|e\rangle$ during its transit through the cavity is proportional to the coupling $g$ as well as the detuning $\Delta$. We assume that one of these two is the parameter to be estimated and for simplifying the following discussion we assume that this parameter is $g$. Once the transit time of the atoms through the cavity is fixed, counting the number of atoms that emerge from the cavity in the states $|e\rangle$ and $|f\rangle$ respectively will yield an estimate of the value of $g$. This estimate is, in practice, obtained by performing a straightforward one parameter fit of the observed statistics to the probability of de-excitation of the atom obtained from the atom-cavity interaction model. Let us assume that atomic transitions out of the state $|e\rangle$ to levels other than $|f\rangle$ due to various reasons is the main noise in the system. In order to estimate $g$ in such a scenario with losses, a model that takes into account the relevant noise processes is required. The dissipation operator $L_e$ in introduced to account for this noise. The action of the jump operator, $L_e$,  takes the atom-cavity system from the state $|e\rangle$ to a state $|s\rangle$ which serves as a placeholder for all other atomic states except $|e\rangle$ and $|f\rangle$.

To understand the dynamics described by Eqs.~\eqref{hamil10} and \eqref{Lindblad2}, we start by identifying the three relevant basis states in the Hilbert space of the atom-cavity system as,
\begin{equation}
    |s\rangle = \begin{pmatrix} 0\\0 \\1\end{pmatrix}\;\; |f\rangle = \begin{pmatrix} 0\\1 \\0\end{pmatrix}\;\; |e\rangle = \begin{pmatrix} 1\\0 \\0\end{pmatrix}\;\;
\end{equation}

We assume that the probe starts in the atom excited state $|f\rangle$, i.e., $\rho_{\rm ff} (0) =1$ to get the following analytical result for the evolution of the elements of the density matrix of the quantum probe:
\begin{eqnarray}
\label{rho1}
\rho_{\rm ff}(t)&=& \frac{e^{-\gamma_e t/2}}{\alpha^2} \bigg[8g^2+(8g^2-\gamma_e^2)\cos \bigg( \frac{\alpha t}{2} \bigg) \nonumber \\
&& \qquad \qquad \qquad \qquad \qquad  + \; \gamma_e \alpha \sin \bigg( \frac{\alpha t}{2} \bigg) \bigg] \nonumber \\
\rho_{\rm ee}(t)&=& \frac{e^{-\gamma_e t/2}}{\alpha^2} 16 g^2 \sin^2 \bigg(\frac{\alpha t}{4} \bigg)\nonumber \\
\rho_{\rm ss}(t)&=& 1- \frac{e^{-\gamma_e t/2}}{\alpha^2} \bigg[ 16g^2 -\gamma_e^2 \cos \bigg( \frac{\alpha t}{2} \bigg) \nonumber \\
&& \qquad \qquad \qquad \qquad \qquad +\gamma_e \alpha \sin \bigg( \frac{\alpha t}{2} \bigg) \bigg] \nonumber \\
\rho_{\rm fe}(t)&=& \rho_{\rm ef}^* =  2 i g \frac{e^{-\gamma_e t/2}}{\alpha^2} \bigg[ \gamma_e - \gamma_e \cos \bigg( \frac{\alpha t}{2} \bigg)  \nonumber \\
&&  \qquad \qquad \qquad \qquad \qquad + \alpha \sin \bigg( \frac{\alpha t}{2} \bigg) \bigg]  \qquad 
\end{eqnarray}
where $\alpha= \sqrt{16g^2 - \gamma_e^2}$ and all other density matrix components are zero. The expressions for the density matrix elements in Eq.~\eqref{rho1} are written suggestively in terms of the parameter $\alpha$. This is because the point $\alpha=0$ corresponds to an exceptional point when the same system is described using non-Hermitian quantum mechanics (see Sec.~\ref{five}). We note here that the limit $\alpha \rightarrow 0$ is well defined for all the components of the density matrix in Eq.~\eqref{rho1}. We can now compute the QFI for this system with respect to either the coupling $g$ or the detuning $\Delta$ using Eq.~\eqref{fisher1}. The QFI corresponding to an estimate of $g$ for different values of the coupling is given in Fig.~\ref{Lindbladf}. We have also plotted the probabilities for being in different states to understand the correlation of the QFI with the dynamics of the system. We see that the QFI is time-dependent and has a peak at some value in time.  This reflects the presence of the noise affecting the estimating process which degrades the sensitivity of the quantum probe with time. In this case, the quantum state of the probe settles down eventually to the state $|s\rangle$ whose evolution at large values of $t$ is not sensitive to changes in $g$. The low values of QFI for small $t$ is because the probabilities $\rho_{\rm ee}$, $\rho_{\rm ff}$ and $\rho_{\rm ss}$ are all changing slowly with time reducing their sensitivity to small changes in $g$. Due to the combined effect of the time taken initially for the QFI to build up and the loss of sensitivity due to noise at later times the QFI has one or more peaks at intermediate times.

\begin{figure}[!htb]
    \centering
    \resizebox{8.5 cm}{6 cm}{\includegraphics{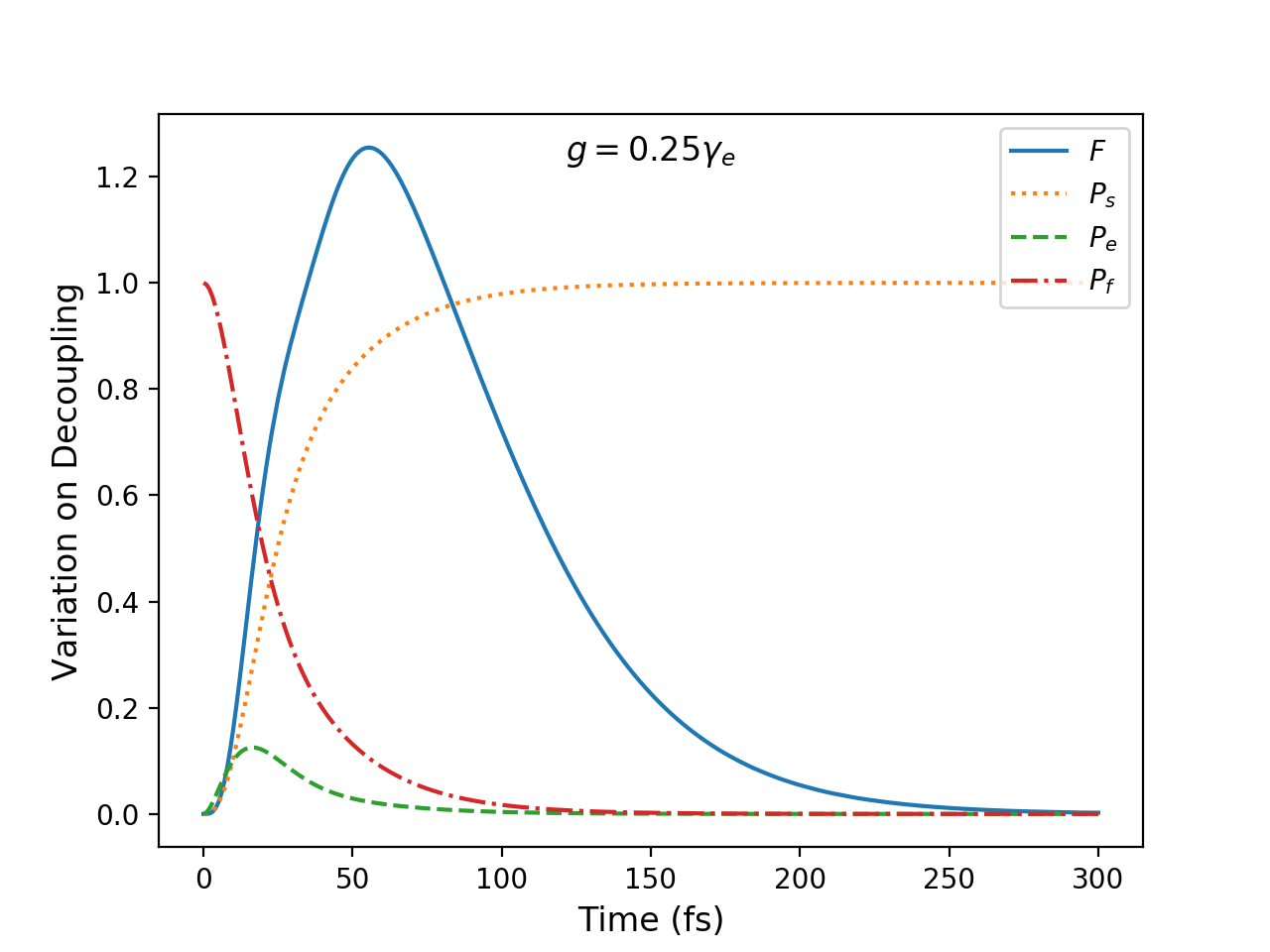}}
    \caption{The dependence of the QFI, $F$ on the detuning parameter $\Delta$ corresponding to the initial state $|f\rangle$ is shown. The parameters used in the plots are: $ \hbar \gamma_e = 0.150 $eV, $g=\gamma_e/4$, $\hbar \Delta=0.02 $eV. Also shown are $\rho_{\rm ee}$, $\rho_{\rm ff}$ and $\rho_{\rm ss}$. Both $F$ and $F_N$ have been scaled down by a factor of 200 to fit in the figure. We see that for estimating $\Delta$, also, $F_{\rm SA}$ serves as a better figure-of-merit for the expected performance of a noisy quantum probe. \label{detuning}}
\end{figure}

Finally we also explore the time dependence of the quantum Fisher information corresponding to the detuning parameter $\Delta$ in Fig.~\ref{detuning}. The system here is initialized in the state $|f\rangle$ with $L_e$ as the collapse operator.

One important point to mention here is that naively it seems that the time dependent quantum fisher information will have a peak where the rate of change of probabilities is the highest. But upon plotting the rate of change of state occupation probabilities one can see that the peak the time dependent QFI occurs at a point in time slightly later that when the rate of change of state occupation probabilities is the highest See Fig.(\ref{probderiv}). 

\begin{figure}[!htb]
    \resizebox{8.5cm}{6 cm}{\includegraphics{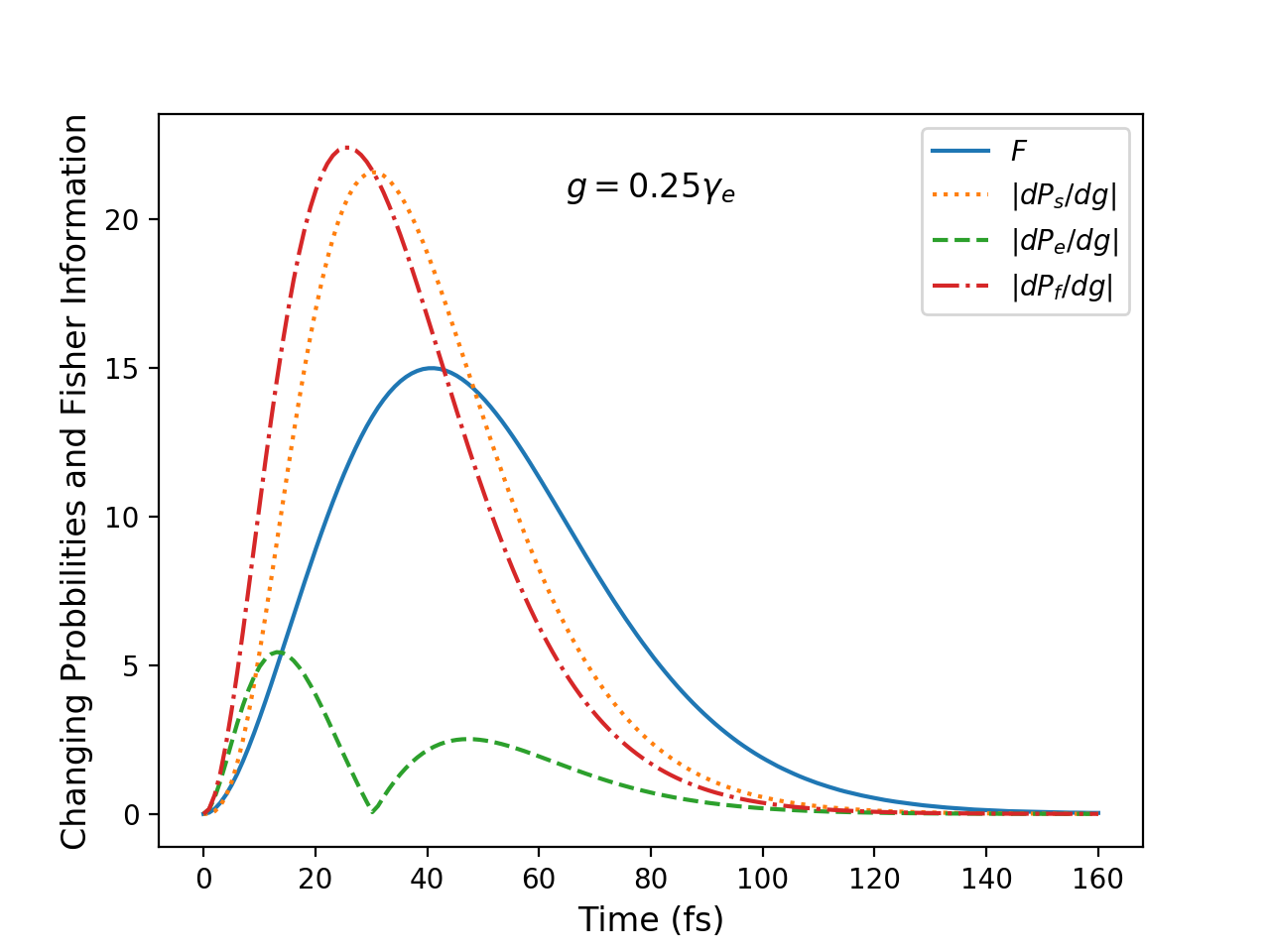}}
    \caption{ The parameters in this figure are the same as the top right subfigure in Fig.~\ref{Lindbladf}. The QFI is plotted along with $|dP_{\rm e}/dg|$ , $|dP_{\rm f}/dg|$ and $|dP_{\rm s}/dg|$ versus time. The QFI has been scaled by a factor of 200}
    \label{probderiv}
\end{figure}

\subsection{Non-Hermitian model}\label{five}

We look at the noisy metrology scheme considered in the previous section again non-Hermitian quantum mechanics. In the non-Hermitian approach, the point $\alpha = 0$ corresponds to an exceptional point at which the non-Hermitian Hamiltonian transitions from being PT-symmetric to not being so. Comparison with the GKSL equation based analysis allows us to see clearly if the exceptional point can lead to advantages in quantum metrology. In the non-Hermitian approach we do not have to consider the sink state and the decay out of the state $|e\rangle$ at a rate $\gamma_e$ can be described directly by adding a term proportional to $i|e\rangle\langle e|$ and modifying the Hamiltonian as
\begin{equation}
    H_{\rm eff} = H - i\hbar \frac{\gamma_e}{2} |e\rangle \langle e| =  H - i\frac{ \hbar}{2} L_e^{\dagger}L_e.
    \label{effectiveH}
\end{equation}
Note that $H$ in Eq.~\eqref{hamil10} is made of only qubit operators and so is $H_{\rm eff}$. We can therefore work entirely in the two dimensional qubit subspace spanned by states $|e\rangle$ and $|f\rangle$ in the non-Hermitian case. Furthermore, we can use state vectors to represent the probe state instead of density matrices and the probe state $|\psi_t\rangle$ evolves as, 
\begin{eqnarray}
    i \hbar \frac{d |\psi_t\rangle}{dt} =  H_{\rm eff}|\psi_t\rangle.
\end{eqnarray}
Heuristically one can understand the connection between the GKSL approach and the non-Hermitian one by noting that in Eq.~\eqref{Lindblad2} the $L_{\rm e} \rho L_{\rm e}^\dagger$ term leads to a contribution proportional to $|s \rangle \langle s|$ in the master equation while the anti-commutator in Eq.~\eqref{Lindblad2} produces contributions proportional to $|e\rangle \langle e|$. The effect of the anti-commutator is captured by the non-Hermitian term that is introduced. A formal development of this connection can be found in~\cite{zloshchastiev_comparison_2014}. 

\begin{figure}[!htb]
    \resizebox{8.5 cm}{6 cm}{\includegraphics{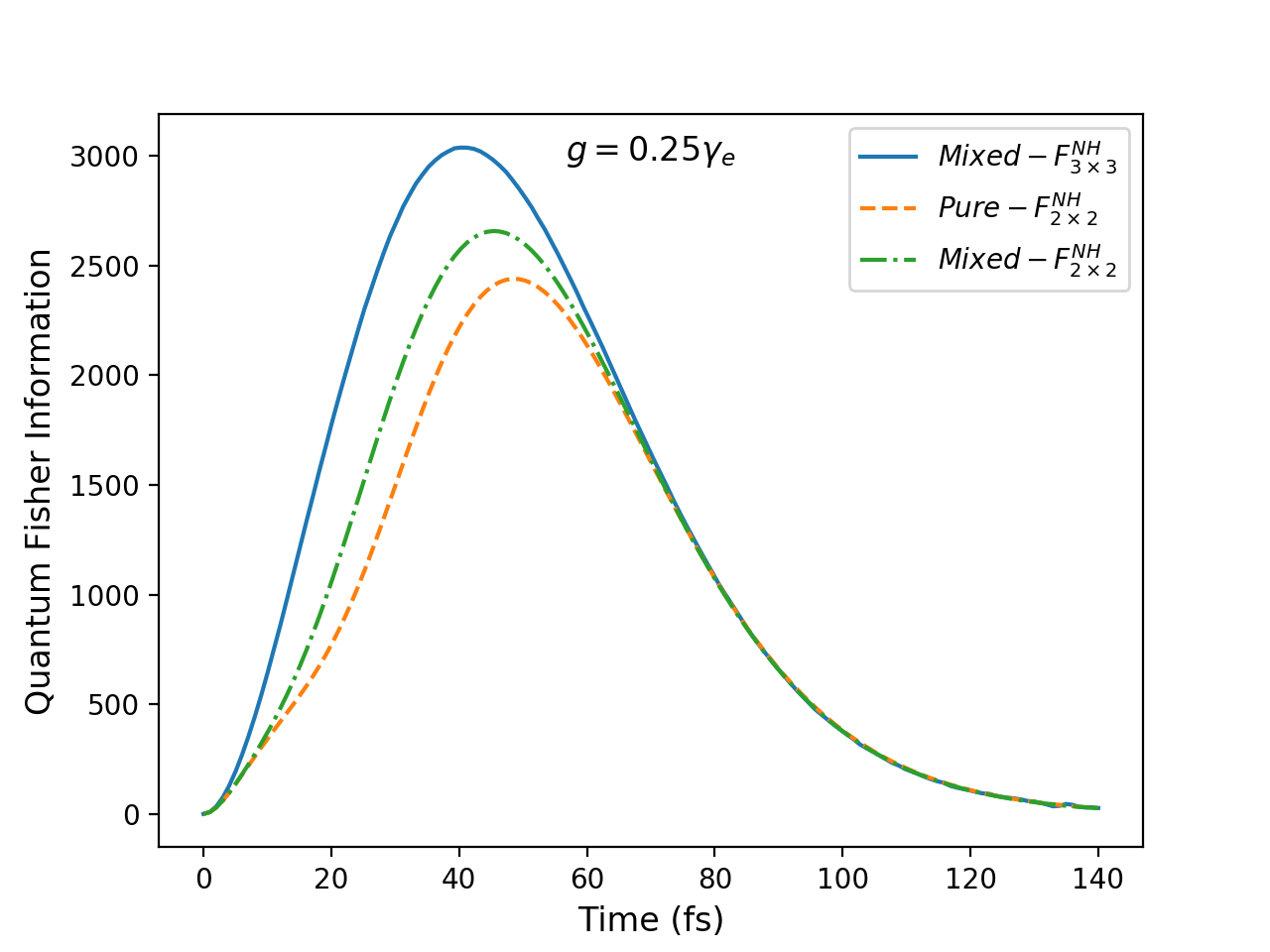}}
    \caption{The plot of QFI, Pure-$F^{NH}_{2\times 2}$ calculated using the pure state vector $|\psi_t^f\rangle$ , the QFI, Mixed-$F^{NH}_{2\times 2}$ calculated using the density matrix $\rho^{NH}_{2\times 2}$ and the QFI, Mixed-$F^{NH}_{3\times 3}$ calculated using the density matrix $\rho^{NH}_{3\times 3}$ vs time is provided. The plot is made at the exceptional point $\alpha=0$.}
    \label{QFIcomparison1}
\end{figure}

We can represent the effective Hamiltonian, $H_{\rm eff}$, by a $2 \times 2$ matrix, 
\begin{equation}
     H_{\rm eff} = \hbar \begin{pmatrix}  \Delta -i\frac{\gamma_{e}}{2} & g\\ g & \Delta \end{pmatrix}
\end{equation}
The eigenvalues and eigenvectors of $H_{\rm eff}$ are,
\begin{equation}
    \lambda_1 = \Delta + \frac{-i \gamma_{e} - \alpha}{4}, \;\;\;\; |\Psi_1\rangle = \frac{1}{\sqrt{2}}\begin{pmatrix}   \frac{-i \gamma_{e} - \alpha}{4g}\\1 \end{pmatrix}
\end{equation}
and 
\begin{equation}
    \lambda_2 = \Delta + \frac{-i \gamma_{e} + \alpha}{4}, \;\;\;\; |\Psi_2\rangle = \frac{1}{\sqrt{2}} \begin{pmatrix}   \frac{-i \gamma_{e} + \alpha}{4g}\\1 \end{pmatrix}
\end{equation}
where, once again we have defined, $\alpha \equiv \sqrt{16g^2- \gamma_{e}^2}$. We can use these eigenvalues and eigenvectors to construct a state $|\psi_t^e \rangle$ which at $t=0$ starts at $|e\rangle$. Such a state is given by, 
\begin{equation}
|\psi_t^e\rangle= \frac{-2g}{\alpha} \left( e^{-i \lambda_1 t}|\Psi_1\rangle- e^{-i \lambda_2 t}|\Psi_2\rangle \right)
\label{estate}
\end{equation}
and $|\psi_0^e\rangle=|e\rangle= ( 1, 0)^T$. We can also construct a state that starts at the initial state $|f\rangle$, 
\begin{equation}
|\psi_t^f\rangle= \frac{8 \sqrt{2} g^2}{\alpha^2 + i \gamma_e\alpha} \left( e^{-i \lambda_1 t}|\Psi_1\rangle + B e^{-i \lambda_2 t}|\Psi_2\rangle \right)
\label{fstate}
\end{equation}
where $B$ is given by, 
\begin{equation}
B= \frac{\alpha^2 - 8 g^2 +i \gamma_e\alpha }{8g^2}
\end{equation}
with $|\psi_0^f\rangle=|f\rangle= ( 0, 1 )^T$. Let us specialize to the case where our initial state is $|f\rangle$. We can use the state $|\psi_t^f\rangle$ to calculate the quantum Fisher information using the pure state formula Eq.~(\ref{pure}). We denote this QFI by Pure-$F^{NH}_{2\times 2}$. This formula however will be incorrect since the derivation of Eq.~(\ref{pure}) assumes the conservation of probabilities and clearly in our non-hermitian case the probabilities are not conserved. We can also use this state to construct the density matrix,
\begin{equation}
    \rho^{\rm NH}_{2\times 2} = |\psi_t^f\rangle \langle \psi_t^f|
\end{equation}
We can use this density matrix in the mixed state QFI formula Eq.~(\ref{fisher1}) to calculate Mixed-$F^{\rm NH}_{2\times 2}$ and as expected we find Pure-$F^{\rm NH}_{2\times 2}\neq$Mixed-$F^{\rm NH}_{2\times 2}$ (see Fig.~\ref{QFIcomparison1}). Let us now extend our $2 \times 2$ desnity matrix to a $3\times 3$ density matrix, 
\begin{equation}
    \rho^{NH}_{3\times3} = \begin{pmatrix} \rho^{\rm NH}_{2\times 2} & 0\\ 0 & \rho_{ss} \end{pmatrix}
    \label{threebythree}
\end{equation}
where $\rho_{\rm ss}= (1- \rho_{\rm ee} - \rho_{\rm ff})$. This density matrix is exactly the density matrix given in Eq.\eqref{rho1} and the QFI calculated with it Mixed-$F^{\rm NH}_{3\times 3}$ is therefore exactly the same as the QFI, $F$ calculated earlier using the GKSL equation (see Fig.~\ref{QFIcomparison1}).  The density matrix $\rho^{\rm NH}_{3\times3}$ has three eigenvalues $\sigma_1, \sigma_2$
and $\sigma_3$ where $\sigma_1$is non zero at all times, $\sigma_2=0$, and $\sigma_3=\rho_{ss}$ due to the block diagonal form of the density matrix. The components of the QFI given in Eq.~\eqref{indicomp} that contribute to the total QFI, $F=$Mixed-$F^{\rm NH}_{3\times 3}$ are $F(\sigma_1,\sigma_1)$, $F(\sigma_1,\sigma_2)$, $F(\sigma_2,\sigma_1)$ and $F(\sigma_3,\sigma_3)$.

An important point to note in Fig.~\ref{QFIcomparison1} is that all the QFI's have been plotted at the exceptional point $\alpha = 0$ and we see no special behavior at that point. This further supports the point that operating a quantum probe at the exceptional point does not yield advantages in quantum metrology as previously pointed out in \cite{Langbein2018,lau2018fundamental}.

\subsection{$N\geq1$ Probes Case}\label{Nscaling}
It is straightforward to generalize the single excitation manifold to the case of $N$ qubits and a resonator mode. See also \cite{cortes2020non}. Our Hamiltonian in this case can be written as, 
\begin{equation}
    H_{\rm eff}^n = H_N - \frac{i \hbar \gamma_e}{2} | e\rangle \langle e| 
    \label{Hamil}
\end{equation}
where 
\begin{equation}
  H_N = \hbar g \sum_{i=1}^{N} \left(| e\rangle \langle f_i| + | f_i\rangle \langle e| \right)  
\end{equation}
and $| e\rangle = (1,0,0 \cdots 0)^T$, $| f_1\rangle = (0,1,0 \cdots 0)^T$, $| f_2\rangle = (0,0,1\cdots 0)^T$ and so on, are $(N+1)$- dimensional column vectors.  We can use the Hamiltonian in Eq.~\eqref{Hamil} to computer the density matrix of the $N$ qubits and the sink state as a function of time as,
\begin{equation}
    \rho^{\rm NH}_{(N+1) \times (N+1)} (t) = \begin{pmatrix} \rho^{\rm NH}_{N \times N} (t) & 0\\ 0 & \rho_{\rm ss}(t) \end{pmatrix}
\end{equation}
where $\rho_{\rm ss}= 1- \rho_{\rm ee}- \sum_i^N \rho_{{\rm f}_i {\rm f}_i}$ as a straightforward generalization of the procedure followed in $N=1$ case. We can then use this density matrix to calculated the QFI for $N$ probes.

\begin{figure}[!htb]
    \resizebox{8 cm}{6 cm}{\includegraphics{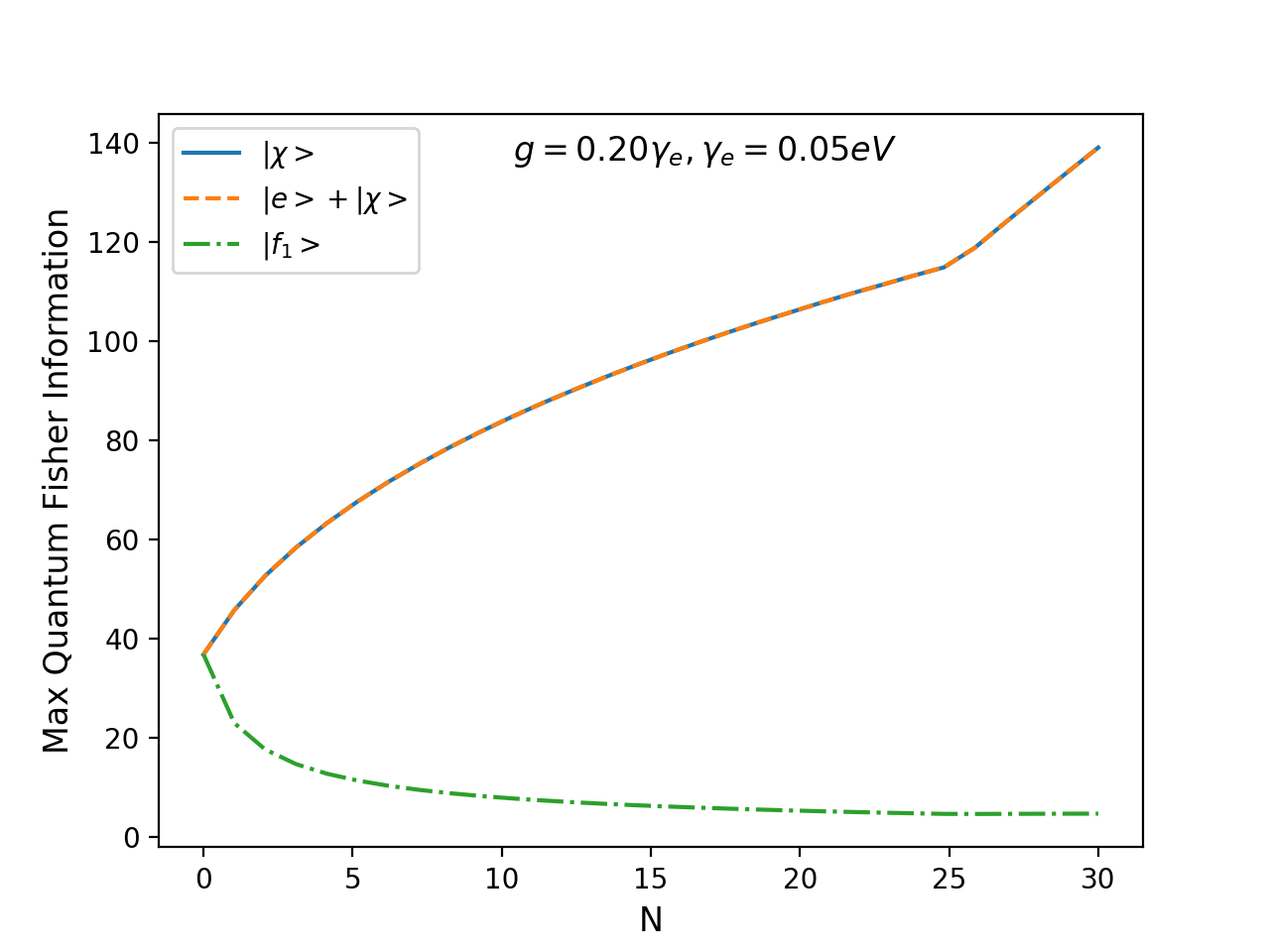}}
    \caption{The maximum value of the quantum Fisher information scaled down by a factor of 1000 is plotted against $N$. We see that the maximum value of the QFI decreases for the state $|f\rangle$ with increasing $N$ and scales as $N^{-0.69}$. The maximum QFI for the initial state $\ket{\chi_1}$ and $|e\rangle + \ket{\chi_1}$  increases with $N$ and scales as $N^{0.52}$ till $N=25$ after which it achieves the SQL and becomes linear with $N$.  }
    \label{MaxQFI1}
\end{figure}

It is then of interest to examine how the QFI scales with increasing $N$ since we know that in the standard quantum limit (SQL) the QFI should increase linearly with $N$ and in the ultimate Heisenberg limit it should scale quadratically with $N$~\cite{boixo_generalized_2007}. Heisenberg scaling can be achieved with certain, highly entangled $N$-qubit states in the noise-free limit \cite{lloyd2004}.
We study the scaling of the maximum value of the QFI for three different initial states, $|e \rangle + \ket{\chi}$, $\ket{\chi}$ and $|f_1\rangle$ in Fig.~\ref{MaxQFI1}. We see that for the  states $|e \rangle + \ket{\chi}$ and $\ket{\chi}$ the peak of the QFI does increase as one increases the $N$. However, as noted in the figure caption, these increases appear to be sublinear, i.e., does not even reach the SQL.
We also see from Fig.~\ref{MaxQFI1} that, for the $|f_1\rangle$ state,  the QFI actually decreases with increasing $N$. Thus for the three particular initial states examined we see that the effect of loss appears to be detrimental to even achieving the SQL, let alone the Heisenberg limit. This seems to be a feature of the fact the we are in 1-excitation manifold and the initial states we can have are the ones with very small amount of entanglement. 

Above we studied the scaling of the QFI numerically but for some special cases some analytical insight can also be obtained. Consider transforming to the basis composed of the states $|e\rangle$,
\begin{equation}
\ket{\chi_1} = \frac{1}{\sqrt{N}} \sum_i \ket{f_i}~~~,
\end{equation}
and $N-1$ states $|\chi_2\rangle$,$|\chi_3\rangle$ $\cdots$ $\ket{\chi_N}$ that are orthogonal to $|e\rangle$ and $|\chi_1\rangle$ for the $N$-qubit Hamiltonian, Eq.~\eqref{Hamil}. In this new basis the dynamics with the single excitation manifold is fully described by a $2\times 2$ sub-block of the Hamiltonian as shown in the appendix, Eq.~\eqref{ham}. See also Ref.~\cite{Campos}. We can see that in this effective two dimensional sub-space the coupling scales as $\sqrt{N}g$. This shows the equivalence of the $N$-qubit model with coupling $g$ to an $N=1$ model with coupling $\sqrt{N} g$. Of course this is not surprising because our model is a variation on the Tavis-Cummings model~\cite{Tavis, Garraway}. Therefore if we choose the initial state to be $|\chi_1 \rangle$ we obtain the following relationship between the QFI's of the $N$-qubit model and the single qubit model,
\begin{equation}
     F( N, g,  \chi_1) = N F(1, \sqrt{N} g,  \chi_1)~~~.
     \label{scale1}
\end{equation}
The origin of the pre-factor $N$ in the above equation is the square of the derivative with respect to $g$ of the density matrix coming from the QFI formula. If, on the other hand, we choose the initial state to be $|f_1 \rangle$, the norm of this state is equivalent to $1/N$ which shows up in the density matrix and cancels the pre-factor in the above equation and we get,
\begin{equation}
     F( N, g, f_1) =  F(1, \sqrt{N} g, f_1)~~~.
     \label{scale2}
\end{equation}
We can easily see from Fig.~{\ref{Lindbladf}} that the QFI for $N=1$ decreases with increasing $g$ and since in the above formula the effective coupling is $\sqrt{N}g$, the function $F(1, \sqrt{N} g)$ will decrease with increasing $N$. 
In Fig.~\ref{MaxQFI1} we also see that after $N=25$ the maximum QFI becomes a linear function of $N$ for the initial states $\ket{\chi_1}$ and $|e\rangle + \ket{\chi_1}$ and we achieve the SQL limit. When $N$ becomes larger, the effective coupling $\sqrt{N}g$ appearing in Eq.~\eqref{scale1} grows while the rate of decay $\gamma_e$ within the single excitation manifold remains unchanged. For large values of the effective coupling, losses are therefore practically negligible and we get back the noiseless case. This is reflected in the second and subsequent peaks of the QFI growing and becoming larger than the first peak indicating that information about the measured parameter continues to be available even after the qubit populations have gone through several rounds of oscillations between the ground state and singly excited state. Both Eqs.~(\ref{scale1}) and (\ref{scale2}), as well as an associated one for starting with all basis components excited are derived in the Appendix.

\section{Time-Dependent Propagation of Error\label{expt}}
The quantum Cramér-Rao bound tells us that $(\delta g(t))^2 \geq 1/{F}$ where $F$ is the quantum Fisher information. We can however also make an estimate  $\delta g(t)$ via the propagation of error formula. 
In this section we will compare the time-dependent QFI with the time-dependent propagation of errors in the estimate of the parameter and see if the minimum value for $\delta g(t)$ occurs at the same point as the minimum in the inverse of the square root of the time dependent QFI $1/\sqrt{F}$. Consider an experiment designed to estimate the parameter $g$. Atoms initialized in the state $|f\rangle$ pass through the cavity and then the number of atoms either in the state $|e\rangle$ or $|f\rangle$ is ascertained. Assuming the initial state is in $|f\rangle$ and the readout on the quantum probe is effected by the projection operator $\Pi_{\rm f} = |f\rangle \langle f|$. Readout of the probe leads to a time dependent signal $\langle \Pi_{\rm f} \rangle_g = \rho_{\rm ff}(t)$, where $\rho_{\rm ff}(t)$ is given in Eq.~\eqref{rho1}. The measured signal is a function of $g$ and using $\langle \Pi_{\rm e} \rangle$ as the estimator for $g$, the propagation of errors formula for the error in the estimate is \cite{kok2020} 
\begin{equation}
\delta g(t) = \frac{\Delta \Pi_{\rm f}(t) }{\big| d\langle \Pi_{\rm f} \rangle_g/dg \big|}~~~, 
\end{equation}
with $\Delta \Pi_{\rm f} = [\langle \Pi_{\rm f}^2 \rangle_g  - \langle \Pi_{\rm f} \rangle_g^2 ]^{1/2}$. Since $\Pi_{\rm f}^2 = \Pi_{\rm f}$, we have
\begin{equation}
\Delta \Pi_{\rm f}(t) =  \rho_{\rm ff}(t)^{1/2}(1-\rho_{\rm ff}(t))^{1/2}. 
\end{equation}
For the case 
$g$ = $\gamma_e/4$ or
$\alpha \rightarrow 0$ we obtain 
\begin{equation}
    \label{eq:measuredvar}
    (\delta g(t))^2 = \frac{36(16 e^{\gamma_e t/2}-(4+\gamma_e t)^2) } {\gamma_e^2 t^4 + (12+\gamma_e t)^2)}
\end{equation}
We see upon plotting in Fig.(\ref{probderiv}) that the minimum in $\delta g(t)$ which was obtained through the error propagation formula aligns with the minimum of $1/\sqrt{F}$. Thus the minimum propagation of errors as a function of time occurs when $F$ is a maximum in time. It should be noted that the QFI result represents an optimization over all possible positive operator values measures (POVM's) and that the propagation of error formula is assuming a particular POVM, $\Pi_f$. Agreement of the time-dependent $\delta g$ from the propagation of error formula and the QFI formula suggests that $\Pi_f$ is nearly optimal. 
\begin{figure}[!htb]
    \resizebox{8.5cm}{6cm}{\includegraphics{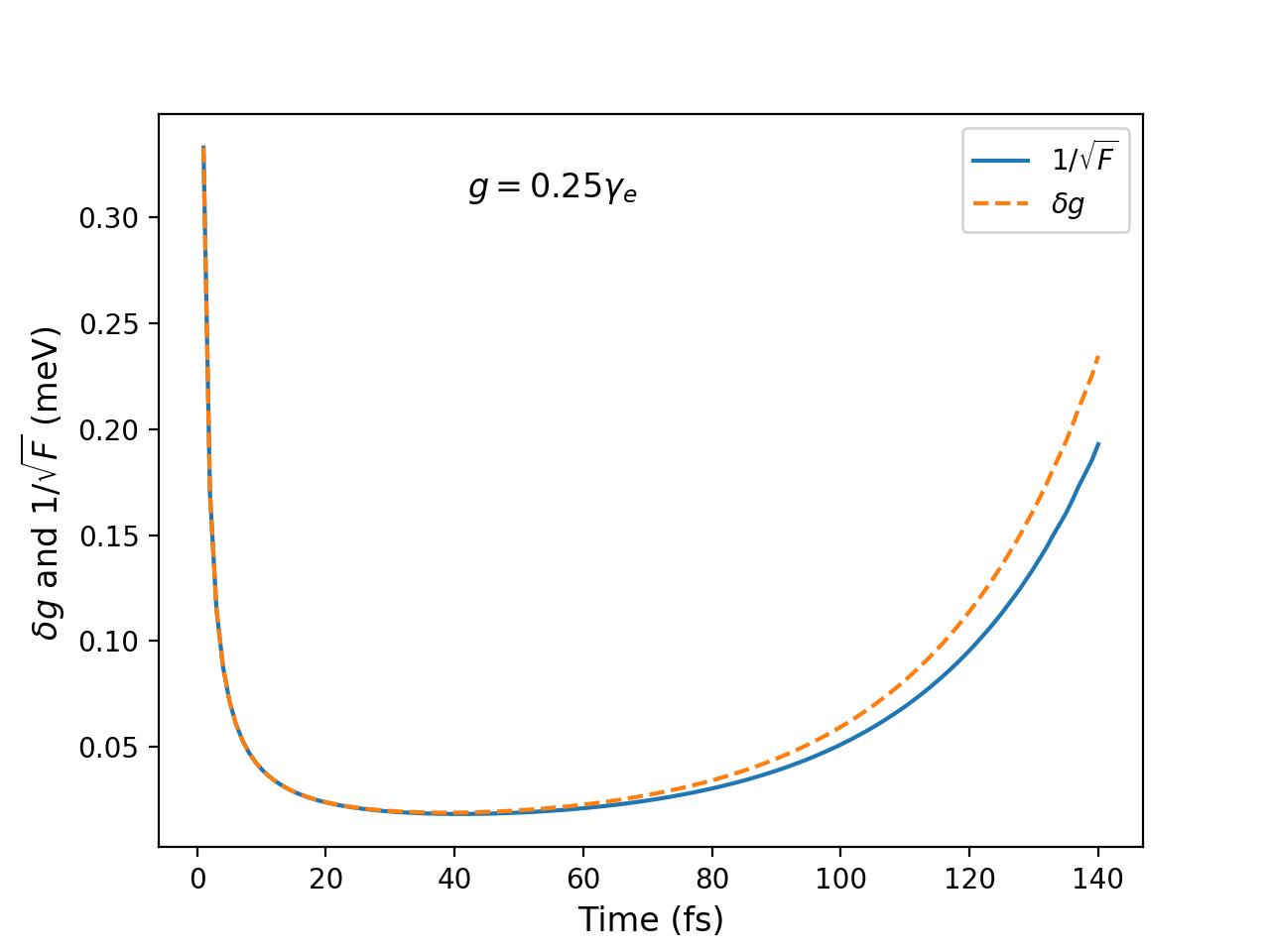}}
    \caption{$1/\sqrt{F}$ and $\delta g$ as functions of time. One can see the that error in the estimate of $g$ obtained via the error propagation formula is the lowest when the QFI is the highest in time. }
    \label{probderiv}
\end{figure} 
We have performed the same calculation when the measured parameter is $\Delta$, the detuning, with results displayed in Fig. \ref{probderiv1}.  Again we see that the minimum of the error in detuning, $\delta \Delta$,  obtained via the error propagation formula coincides with the minimum in $1/\sqrt{F}$.

\begin{figure}[!htb]
    \resizebox{8.5cm}{6 cm}{\includegraphics{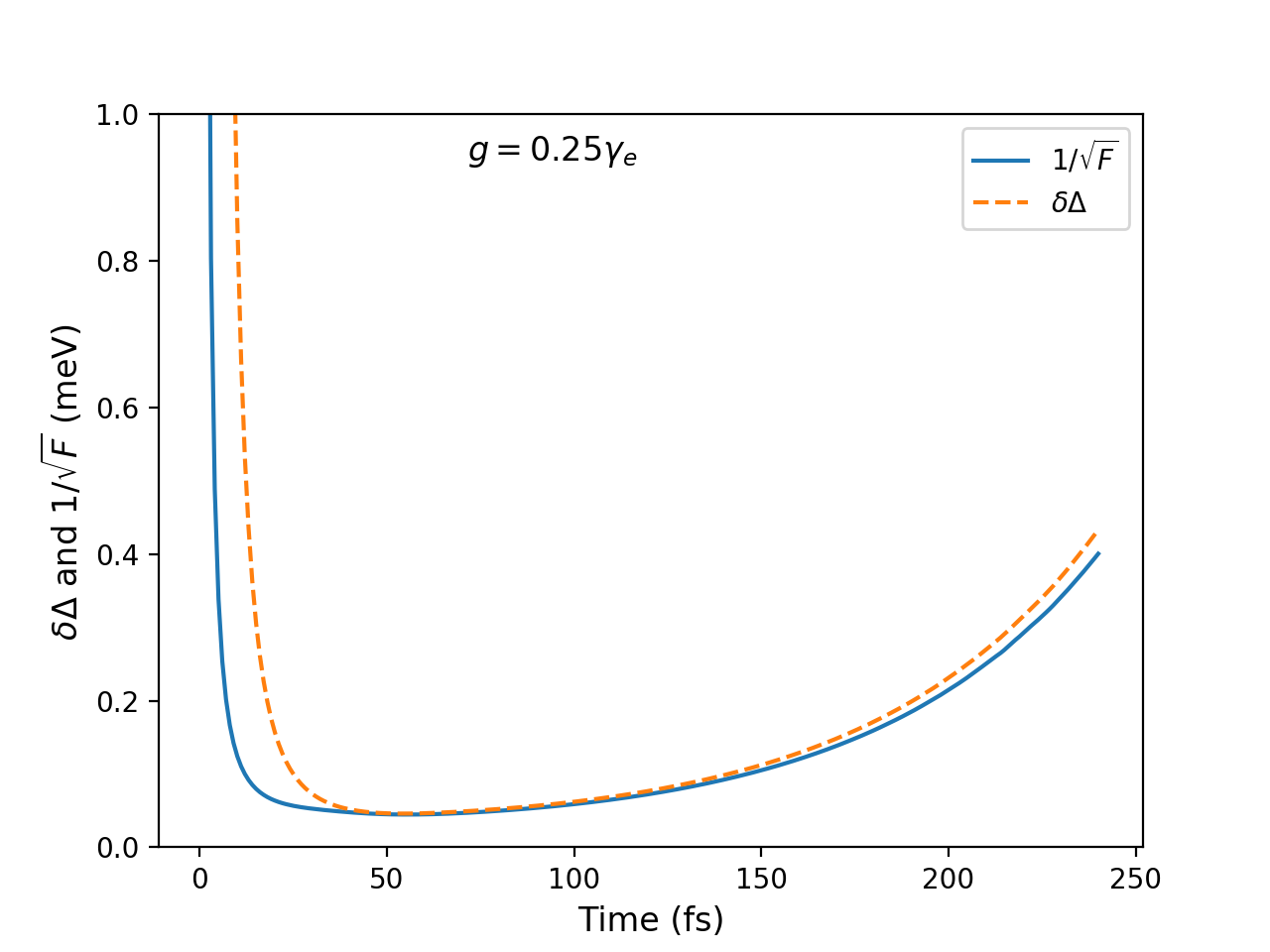}}
    \caption{$1/\sqrt{F}$ and $\delta \Delta$ as functions of time. Just like the parameter estimation of the coupling $g$, the error in the estimate of $\Delta$ obtained via the error propagation formula is the lowest when the QFI is the highest in time.  }
    \label{probderiv1}
\end{figure} 



\section{A Procedure for Parameter Estimation\label{optimal}}
Here we describe a possible experimental procedure for parameter estimation making use of the optimum time concept in open quantum systems. 
Some care is needed because the system's detailed time evolution, and thus the optimum time, depends on the parameter(s) (such as $g$ which we focus on here) that one is attempting to infer.
First, like most quantum sensing and metrology cases, one develops a reasonable model for the system. One then carries out low-resource-consuming experiments to get an idea of $g$  and the corresponding optimal time or time range via fitting to the model.  For example, these experiments could be relatively low-shot-count measurements of the relevant probabilities over a wide time range leading to an estimate of $g$, called $g_0$.  The time-dependent QFI can be determined for this value of $g$ and an optimum time range centered around the optimal time identified. The optimal time is the point in time where the time dependent QFI is the largest. Calculations for a range of $g$ near $g_0$ may also be carried out to obtain an idea of the sensitivity of the optimum time range to $g$ variations. This is also why not a single time but a range of times where the QFI is large (for possible variations in $g$) is probably best. Next, one invests more significant experimental resources (e.g., many more shots), focusing on fitting the model’s $g$ parameter to data in just the optimum time range to obtain an improved, more precise value of $g$. 

\begin{figure}[!htb]
    \resizebox{8.5cm}{6 cm}{\includegraphics{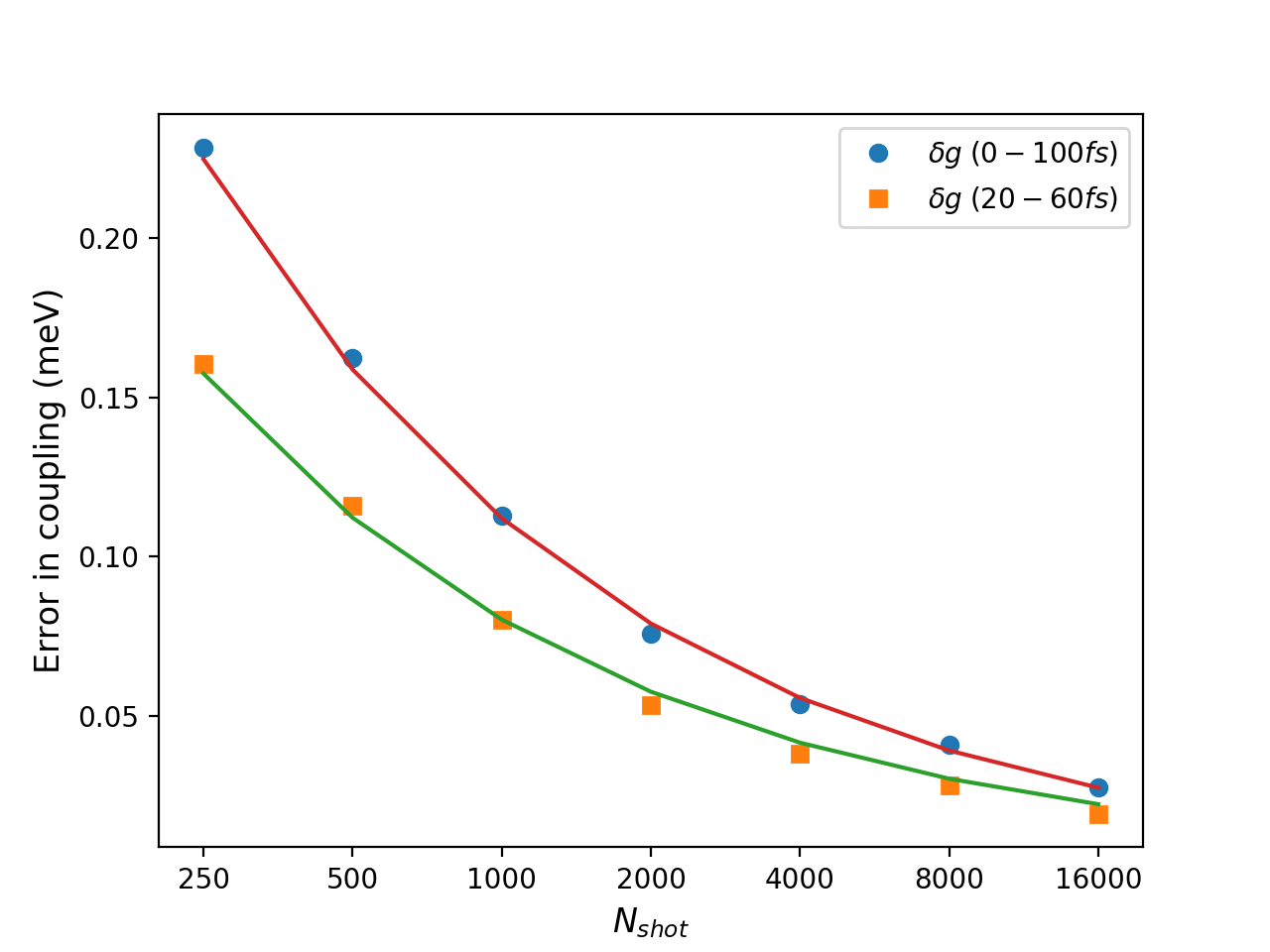}}
    \caption{The error for the estimate of $g$ when performing the experiments in the full time range (0-100fs) (blue circles) vs the error when performing the measurement in the time range (20-60fs) centered around the optimal point in time approx 40fs (orange squares). The smooth curves are fits to $a/\sqrt{N_{shot}}$. }
    \label{shotnoise}
\end{figure}

To demonstrate the viability of the approach suggested above, we applied our procedure using simulated noisy data for the single probe case (Fig. 1) with $g$ = 0.25 $\gamma_e$. First, we consider  fifty evenly spaced times across a large (0, 100 fs) time range and imagine measuring noisy probabilities for being in state $f$ for each time via repeated shots. Since we know the underlying true mean as a function of time, $P_f(t)=\rho_{ff}(t)$ from Eq.~\eqref{rho1}, a random draw from a binomial distribution  $Pr(N_f(t); N_{shot},P_f(t))$ will give an integer, $N_f(t)$,  between 0 and $N_{shot}$ that represents a possible number of successful observations of state $f$ out of $N_{shot}$ tries;  the noisy probability for that time is $N_f(t)/N_{shot}$. (Rather than use just one draw we generally average over a number of experiments, $M$, for each $N_{shot}$ case, with $M$ = 100.) We fit the model, $P_f(t)$ to the fifty noisy time samples to obtain an estimate of $g$ for that given $N_{shot}$ and repeat the procedure for a variety of $N_{shot}$ values. Figure \ref{shotnoise} displays the error in this procedure as a function of $N_{shot}$, i.e. the square root of the root mean square error or variance of the fitted values of $g$ from the known true value.  Of course this is a quantity not available in a real experiment where the precise value of $g$ is not known, but theoretically it connects with the uncertainty of the estimator.  The orange curve is a fit to $a/\sqrt{N_{shot}}$ and we see that, as might be expected, the error scales as $1/\sqrt{N_{shot}}$ and can be driven down to very small values with ever increasing $N_{shot}$.  Now it turns out that (not knowing the exact $g$ or this estimator error) low-resource-consuming experiments, e.g., $N_{shot} \leq 500$ in Fig. \ref{shotnoise}, can reasonably estimate $g$ to 1-10 \%.  If we call such a value $g_0$, we can evaluate the QFI as a function of time using our model and identify an optimum time region. We note (Fig. 1) that the QFI is relatively broad and that additional calculations with the model can be made to ascertain the sensitivity of the optimal time range to variations in $g$.  There may be problems different from this one where there is too much sensitivity to $g$ and this procedure  will, at least, rule those out as candidates for our scheme.

Suppose now that one has identified the optimum time and selected a range about it.  In our example, it turns out the time is approximately 40 fs and we have chosen the 20-60 fs region.  We repeat the shot noise simulated experiments as described above, but instead of fitting to time data in the entire 0-100 fs region, we fit the model to time data in just the 20 - 60 fs region. The resulting parameter errors as a function of $N_{shot}$ are the orange squares in Fig. \ref{shotnoise} and the green curve is a fit of those points to $a/\sqrt{N_{shot}}$.  Notice that for the same amount of resources consumed 
(i.e., same value of $N_{shot}$) the parameter error is always less than that from fitting over the entire time range. To achieve comparable error to the optimum time range results for a given $N_{shot}$, the full time range results would have to be run with about twice the resources, i.e. for 2$N_{shot}$ shots.  Thus if the first low-resource step in our procedure can be carried out employing  less than  $N_{shot}$ shots, then the second step here will achieve, for less total resources, the desired estimate of the parameter for a total resource cost of less than 2$N_{shot}$. 

We note that we have also carried out the procedure in the above paragraph but with sub-optimal time ranges in terms of the QFI magnitude  such as 0 - 40 fs.  The result (not shown) yields errors significantly larger than those for the optimum time range, consistent with our expectations.

We stress again, of course, that in these simulations we know what the exact value of $g$ is and so can calculate the actual errors that are displayed in Fig. \ref{shotnoise}. However, having possession of a good underlying parameterized model, a requirement for most metrology and sensing scenarios, and the ability to calculate the time-dependent QFI of the model, can allow one to find a likely time region where the parameter can be most efficiently estimated and then focus resources on that region.


\section{Conclusions and Future Directions}\label{six}

The quantum Fisher information is an important tool for characterizing quantum sensing and metrology and it bounds the accuracy with which a parameter can be estimated using a in measurement scheme that uses quantum probes. In this work we studied the time-dependent quantum Fisher information for an atom-cavity open quantum system and focused on estimating the coupling $g$ of the atom with the cavity and the detuning $\Delta$. Due to there being losses in the system we find that the time-dependent quantum Fisher information has a peak value in time, suggesting that there is an optimal point in time where the quantum measurements should be performed to obtain the best possible estimate of $g$ and $\Delta$. We use the non-hermitian formalism to extend our results to the case when $N$-probes are used for quantum sensing. These results are then used to study the scaling of the maximum value of the time dependent QFI with $N$ number of atoms in our quantum system when limited to the manifold of singly excited states. Here we find that the losses have a very detrimental effect and does not allow the measurement to even achieve the standard quantum limit (SQL).

We have compared our calculation of the time dependent QFI with the propagation of error in the variance of the parameter to be estimated. Our calculations show that when the optimal estimator is used the the minimum value of the error in the estimate of the parameter in time obtained from the propagation of error calculation coincides with the point in time where the QFI is the highest. Lastly we have provided a procedure that allows us to use the optimal time concept to improve the accuracy of our estimate of the unknown parameter given a finite amount of quantum resources available for performing the quantum measurements. 

In our future work we plan to apply the optimal time concept for performing quantum measurements to other open quantum systems. We will also study the scaling of the maximum value of the time-dependent QFI in the presence of dissipation when initial states with more entanglement in them (such as GHZ or Dicke states in higher excitation manifolds) are used.

\section*{Acknowledgments}
This material is based upon work supported by the U.S. Department of Energy Office of Science National Quantum Information Science Research Centers. S.K.G. and Z.H.S. were supported by the Q-NEXT Center. Work performed at the Center for Nanoscale Materials, a U.S. Department of Energy Office of Science User Facility, was supported by thee U.S. DOE Office of Basic Energy Sciences, under Contract No. DE-AC02-06CH11357. Z.H.S. would like to thank Michael Perlin for helpful discussions. S.K.G. benefited from disucussions with Cristian L. Cortes. A.~S.~was supported by QuEST grant No Q-113 of the Department of Science and Technology, Government of India. 

\section*{Appendix}\label{appendix}
The matrix representation of our Hamiltonian Eq.~{\eqref{Hamil}} in the $|e\rangle$, $|f_1\rangle$, $|f_2\rangle$, $\cdots$ basis. is given by 
\begin{equation}
    H = \hbar \begin{pmatrix} -i\frac{\gamma_e}{2} & g &g & \dots\\ g & 0 & 0& \dots \\g & 0 & 0 & \dots \\ \vdots & \vdots & & \ddots   \end{pmatrix}
\end{equation}
Let us transform to new orthogonal basis $|e\rangle$, $|\chi_1\rangle$, $|\chi_2\rangle$ $\cdots$, where $|\chi_1\rangle$ is given by 
\begin{equation}
\ket{\chi_1} = \frac{1}{\sqrt{N}} \sum_i \ket{f_i}~~~.
\end{equation}
By construction, $\ket{e}$ and $\ket{\chi_1}$ are orthogonal.  We can find the
remaining $|\chi_2\rangle$,$|\chi_3\rangle$, $\cdots$, $\ket{\chi_N}$  basis states that are orthogonal to $|e\rangle$ and $|\chi_1\rangle$ via a Gram-Schmidt procedure applied to $\ket{f_2}$, $\cdots$, $\ket{f_N}$, for example. In this new basis the matrix representation of our Hamiltonian is given by
\begin{eqnarray}
    H  =  \hbar \begin{pmatrix} H_{2 \times 2} & 0& \dots\\ 0 & 0 \\ \vdots &  & \ddots   \end{pmatrix} ~~, \nonumber \\
    H_{2 \times 2}  = \begin{pmatrix}-i\frac{\gamma_e}{2} & \sqrt{N}g \\ \sqrt{N}g & 0 \end{pmatrix}~~,
    \label{ham}
\end{eqnarray}
Now the Hamiltonian matrix in this basis is effectively a $2\times 2 $ block involving the $|e\rangle$ state and the $|\chi_1 \rangle$ state above, with coupling $\sqrt{N} g$. This shows that the $N$-atom Hamiltonian with coupling $g$ can be essentially described by a $N=1$ Hamiltonian with a $\sqrt{N}g$ coupling between its two states. For a related treatment, see Ref.~\cite{Campos}.


It is now possible to write down exact expressions for the time evolved density matrix and the quantum Fisher information all in terms of the solution of the $2\times 2 $ problem. Given the block diagonal structure of our Hamiltonian, Eq.~(\ref{ham}), our density matrix $\rho(t)$ will also have a block diagonal structure in the new basis, 
\begin{equation}
    \rho(t) = \begin{pmatrix} \rho_{ee} & \rho_{e 1} & 0 & 0& 0&\dots\\  \rho_{ 1e} & \rho_{11} & 0 &0 & 0&\cdots  \\ 0 & 0 & \rho_{22} & 0& 0&\dots \\  0&0&0& \rho_{33} &0& \dots \\ \vdots & \vdots & \vdots& \vdots& \ddots& \dots   \end{pmatrix} ~~~,
    \label{rhoeqn}
\end{equation}
where the indices 1, 2, $\cdots$, $N$ refer to $\ket{\chi_1}$, $\ket{\chi_2}$, $\cdots$, $\ket{\chi_N}$.
A general initial condition in the original basis $|e\rangle, |f_1\rangle, |f_2 \rangle , \cdots$ will have the form, 
\begin{equation}
    \ket{\psi(0)} = a_e |e\rangle + \sum_{k=1}^N a_k |\chi_k \rangle ~~,
\end{equation}
in the $|e\rangle$, $|\chi_1\rangle$, $|\chi_2\rangle$ $\cdots$ basis. The density matrix elements $\rho_{22}, \rho_{33}, \cdots$ will be constant in time. If we define the re-scaled coupling to be $G= \sqrt{N} g$ for the $H_{2\times 2}$, we have $d\rho_{\rm kk}/dG = 0$ for all $k=2,3,\cdots$. For the density matrix elements that are not constant, we have, 
\begin{equation}
    \frac{d \rho}{dg} = \sqrt{N} \frac{d \rho}{dG}~~.
\end{equation}
Since the general form of the quantum Fisher information, Eqs.~(\ref{fisher1}) and (\ref{indicomp}), involves
derivatives with respect to $g$ (or $G$), the quantum Fisher information
inferred from just the upper $2\times 2$ part of Eq.~(\ref{rhoeqn}) (along with
any auxiliary quantities such as $\rho_{\rm ss}$ discussed below) will be the
quantum Fisher information for the full problem.
Since the square of the density matrix with respect to $g$ is what enters the quantum Fisher information formula, 
\[ \bigg(\frac{d \rho}{dg} \bigg)^2 = N \bigg( \frac{d \rho}{dG} \bigg)^2, \] 
it is clear that, for some initial condition $\ket{\psi (0)}$,
\begin{equation}
F( N, g,\psi(0) ) = N \;F(1, G=\sqrt{N} g, \psi (0))~~,
\label{qfiscale}
\end{equation}
This means that the $N$ qubit quantum Fisher information with coupling $g$ can be obtained from an $N$ = 1
qubit quantum Fisher information with coupling $G$ = $\sqrt{N}g$ based on the $2\times 2$, $\ket{e}$ - $\ket{\chi_1}$ subspace and with an appropriate initial condition, $a_e(0), \, a_1(0)$ (inferred from $\psi (0))$. In addition to the factor of $N$ on the right-hand side of Eq.~(\ref{qfiscale}), there can be additional $N$ dependence arising from the initial state employed.
Let us analyze several initial states. 
\begin{itemize}
    \item $\ket{\psi(0)} = \frac{1}{\sqrt{N}}\sum_i |f_i \rangle = |\chi_1 \rangle$:
     In this case we have $a_e= \langle e| \psi (0) \rangle =0$ and $a_1= \langle \chi_1| \psi (0) \rangle =1$ and therefore no extra $N$ dependence is introduced in the Eq.~(\ref{qfiscale}). 
    \item $\ket{\psi(0)} = |f_1\rangle$: For this initial state $a_e= \langle e| \psi (0) \rangle =0$ and $a_1= \langle \chi_1| \psi (0) \rangle =1/\sqrt{N}$. We will therefore have to solve our $2 \times 2$ system with this initial condition which overall has norm 1/$N$ instead of unity.  We could equivalently solve the $2 \times 2$ system with initial condition $a_e$ = 0 and $a_1$ = 1, and multiply the overall result by 1/$N$. In the latter case this factor will cancel the prefactor on the right hand side in the Eq.~(\ref{qfiscale}). 
    
    \item $\ket{\psi(0)} = \frac{1}{\sqrt{N+1}}( |e\rangle + \sum_i |f_i \rangle)$: For this initial condition, $a_e= \langle e| \psi (0) \rangle =1/\sqrt{N+1}$ and $a_1= \langle \chi_1| \psi (0) \rangle =\sqrt{N/(N+1)}$. Notice here that while $\ket{e}$ and all the $\ket{f_i}$ have equal amplitudes in the original basis, $\ket{e}$ and $\ket{a_1}$ are not equally weighted. As with the first example, the pre-factor of $N$ in Eq.~(\ref{qfiscale}) will remain in tact.
\end{itemize}
In the formula for $F(1, G=\sqrt{N} g, \psi(0))$ the density matrix elements $\rho_{ee},\rho_{1e},\rho_{e1}$ and $\rho_{11}$ contribute. However our open quantum system has a sink state as well. The density matrix element $\rho_{ss}$ corresponding to the sink state can be calculated via $\rho_{ss} (t)= 1- \rho_{ee} (t) - \rho_{11} (t) -\rho_{22} (0) -\cdots-\rho_{NN} (0) $. We will therefore have to lift our $2 \times 2$ density matrix to the $3 \times 3$ density matrix as given in Eq.~(\ref{threebythree}) to calculate $F(1, G=\sqrt{N} g , \psi(0))$.

\bibliography{references}

\begin{thebibliography}{34}%
\makeatletter
\providecommand \@ifxundefined [1]{%
 \@ifx{#1\undefined}
}%
\providecommand \@ifnum [1]{%
 \ifnum #1\expandafter \@firstoftwo
 \else \expandafter \@secondoftwo
 \fi
}%
\providecommand \@ifx [1]{%
 \ifx #1\expandafter \@firstoftwo
 \else \expandafter \@secondoftwo
 \fi
}%
\providecommand \natexlab [1]{#1}%
\providecommand \enquote  [1]{``#1''}%
\providecommand \bibnamefont  [1]{#1}%
\providecommand \bibfnamefont [1]{#1}%
\providecommand \citenamefont [1]{#1}%
\providecommand \href@noop [0]{\@secondoftwo}%
\providecommand \href [0]{\begingroup \@sanitize@url \@href}%
\providecommand \@href[1]{\@@startlink{#1}\@@href}%
\providecommand \@@href[1]{\endgroup#1\@@endlink}%
\providecommand \@sanitize@url [0]{\catcode `\\12\catcode `\$12\catcode
  `\&12\catcode `\#12\catcode `\^12\catcode `\_12\catcode `\%12\relax}%
\providecommand \@@startlink[1]{}%
\providecommand \@@endlink[0]{}%
\providecommand \url  [0]{\begingroup\@sanitize@url \@url }%
\providecommand \@url [1]{\endgroup\@href {#1}{\urlprefix }}%
\providecommand \urlprefix  [0]{URL }%
\providecommand \Eprint [0]{\href }%
\providecommand \doibase [0]{https://doi.org/}%
\providecommand \selectlanguage [0]{\@gobble}%
\providecommand \bibinfo  [0]{\@secondoftwo}%
\providecommand \bibfield  [0]{\@secondoftwo}%
\providecommand \translation [1]{[#1]}%
\providecommand \BibitemOpen [0]{}%
\providecommand \bibitemStop [0]{}%
\providecommand \bibitemNoStop [0]{.\EOS\space}%
\providecommand \EOS [0]{\spacefactor3000\relax}%
\providecommand \BibitemShut  [1]{\csname bibitem#1\endcsname}%
\let\auto@bib@innerbib\@empty
\bibitem [{\citenamefont {Braunstein}\ and\ \citenamefont
  {Caves}(1994)}]{braunstein1994statistical}%
  \BibitemOpen
  \bibfield  {author} {\bibinfo {author} {\bibfnamefont {S.~L.}\ \bibnamefont
  {Braunstein}}\ and\ \bibinfo {author} {\bibfnamefont {C.~M.}\ \bibnamefont
  {Caves}},\ }\bibfield  {title} {\bibinfo {title} {Statistical distance and
  the geometry of quantum states},\ }\href@noop {} {\bibfield  {journal}
  {\bibinfo  {journal} {Physical {R}eview {L}etters}\ }\textbf {\bibinfo
  {volume} {72}},\ \bibinfo {pages} {3439} (\bibinfo {year}
  {1994})}\BibitemShut {NoStop}%
\bibitem [{\citenamefont {Giovannetti}\ \emph
  {et~al.}(2004{\natexlab{a}})\citenamefont {Giovannetti}, \citenamefont
  {Lloyd},\ and\ \citenamefont {Maccone}}]{giovannetti_quantum-enhanced_2004}%
  \BibitemOpen
  \bibfield  {author} {\bibinfo {author} {\bibfnamefont {V.}~\bibnamefont
  {Giovannetti}}, \bibinfo {author} {\bibfnamefont {S.}~\bibnamefont {Lloyd}},\
  and\ \bibinfo {author} {\bibfnamefont {L.}~\bibnamefont {Maccone}},\
  }\bibfield  {title} {\bibinfo {title} {Quantum-{Enhanced} {Measurements}:
  {Beating} the {Standard} {Quantum} {Limit}},\ }\href
  {https://doi.org/10.1126/science.1104149} {\bibfield  {journal} {\bibinfo
  {journal} {Science}\ }\textbf {\bibinfo {volume} {306}},\ \bibinfo {pages}
  {1330} (\bibinfo {year} {2004}{\natexlab{a}})}\BibitemShut {NoStop}%
\bibitem [{\citenamefont {Degen}\ \emph {et~al.}(2017)\citenamefont {Degen},
  \citenamefont {Reinhard},\ and\ \citenamefont
  {Cappellaro}}]{degen2017quantum}%
  \BibitemOpen
  \bibfield  {author} {\bibinfo {author} {\bibfnamefont {C.~L.}\ \bibnamefont
  {Degen}}, \bibinfo {author} {\bibfnamefont {F.}~\bibnamefont {Reinhard}},\
  and\ \bibinfo {author} {\bibfnamefont {P.}~\bibnamefont {Cappellaro}},\
  }\bibfield  {title} {\bibinfo {title} {Quantum sensing},\ }\href@noop {}
  {\bibfield  {journal} {\bibinfo  {journal} {Reviews of {M}odern {P}hysics}\
  }\textbf {\bibinfo {volume} {89}},\ \bibinfo {pages} {035002} (\bibinfo
  {year} {2017})}\BibitemShut {NoStop}%
\bibitem [{\citenamefont {Braun}\ \emph {et~al.}(2018)\citenamefont {Braun},
  \citenamefont {Adesso}, \citenamefont {Benatti}, \citenamefont {Floreanini},
  \citenamefont {Marzolino}, \citenamefont {Mitchell},\ and\ \citenamefont
  {Pirandola}}]{braun_quantum-enhanced_2018}%
  \BibitemOpen
  \bibfield  {author} {\bibinfo {author} {\bibfnamefont {D.}~\bibnamefont
  {Braun}}, \bibinfo {author} {\bibfnamefont {G.}~\bibnamefont {Adesso}},
  \bibinfo {author} {\bibfnamefont {F.}~\bibnamefont {Benatti}}, \bibinfo
  {author} {\bibfnamefont {R.}~\bibnamefont {Floreanini}}, \bibinfo {author}
  {\bibfnamefont {U.}~\bibnamefont {Marzolino}}, \bibinfo {author}
  {\bibfnamefont {M.~W.}\ \bibnamefont {Mitchell}},\ and\ \bibinfo {author}
  {\bibfnamefont {S.}~\bibnamefont {Pirandola}},\ }\bibfield  {title} {\bibinfo
  {title} {Quantum-enhanced measurements without entanglement},\ }\href
  {https://doi.org/10.1103/RevModPhys.90.035006} {\bibfield  {journal}
  {\bibinfo  {journal} {Reviews of Modern Physics}\ }\textbf {\bibinfo {volume}
  {90}},\ \bibinfo {pages} {035006} (\bibinfo {year} {2018})}\BibitemShut
  {NoStop}%
\bibitem [{\citenamefont {Pezzè}\ \emph {et~al.}(2018)\citenamefont {Pezzè},
  \citenamefont {Smerzi}, \citenamefont {Oberthaler}, \citenamefont {Schmied},\
  and\ \citenamefont {Treutlein}}]{pezze_quantum_2018}%
  \BibitemOpen
  \bibfield  {author} {\bibinfo {author} {\bibfnamefont {L.}~\bibnamefont
  {Pezzè}}, \bibinfo {author} {\bibfnamefont {A.}~\bibnamefont {Smerzi}},
  \bibinfo {author} {\bibfnamefont {M.~K.}\ \bibnamefont {Oberthaler}},
  \bibinfo {author} {\bibfnamefont {R.}~\bibnamefont {Schmied}},\ and\ \bibinfo
  {author} {\bibfnamefont {P.}~\bibnamefont {Treutlein}},\ }\bibfield  {title}
  {\bibinfo {title} {Quantum metrology with nonclassical states of atomic
  ensembles},\ }\href {https://doi.org/10.1103/RevModPhys.90.035005} {\bibfield
   {journal} {\bibinfo  {journal} {Reviews of Modern Physics}\ }\textbf
  {\bibinfo {volume} {90}},\ \bibinfo {pages} {035005} (\bibinfo {year}
  {2018})},\ \bibinfo {note} {publisher: American Physical Society}\BibitemShut
  {NoStop}%
\bibitem [{\citenamefont {Giovannetti}\ \emph {et~al.}(2006)\citenamefont
  {Giovannetti}, \citenamefont {Lloyd},\ and\ \citenamefont
  {Maccone}}]{giovannetti2006quantum}%
  \BibitemOpen
  \bibfield  {author} {\bibinfo {author} {\bibfnamefont {V.}~\bibnamefont
  {Giovannetti}}, \bibinfo {author} {\bibfnamefont {S.}~\bibnamefont {Lloyd}},\
  and\ \bibinfo {author} {\bibfnamefont {L.}~\bibnamefont {Maccone}},\
  }\bibfield  {title} {\bibinfo {title} {Quantum metrology},\ }\href@noop {}
  {\bibfield  {journal} {\bibinfo  {journal} {Physical {R}eview {L}etters}\
  }\textbf {\bibinfo {volume} {96}},\ \bibinfo {pages} {010401} (\bibinfo
  {year} {2006})}\BibitemShut {NoStop}%
\bibitem [{\citenamefont {Giovannetti}\ \emph {et~al.}(2011)\citenamefont
  {Giovannetti}, \citenamefont {Lloyd},\ and\ \citenamefont
  {Maccone}}]{giovannetti_advances_2011}%
  \BibitemOpen
  \bibfield  {author} {\bibinfo {author} {\bibfnamefont {V.}~\bibnamefont
  {Giovannetti}}, \bibinfo {author} {\bibfnamefont {S.}~\bibnamefont {Lloyd}},\
  and\ \bibinfo {author} {\bibfnamefont {L.}~\bibnamefont {Maccone}},\
  }\bibfield  {title} {\bibinfo {title} {Advances in quantum metrology},\
  }\href {https://doi.org/10.1038/nphoton.2011.35} {\bibfield  {journal}
  {\bibinfo  {journal} {Nature Photonics}\ }\textbf {\bibinfo {volume} {5}},\
  \bibinfo {pages} {222} (\bibinfo {year} {2011})}\BibitemShut {NoStop}%
\bibitem [{\citenamefont {Barbieri}(2022)}]{barbieri_optical_2022}%
  \BibitemOpen
  \bibfield  {author} {\bibinfo {author} {\bibfnamefont {M.}~\bibnamefont
  {Barbieri}},\ }\bibfield  {title} {\bibinfo {title} {Optical {Quantum}
  {Metrology}},\ }\href {https://doi.org/10.1103/PRXQuantum.3.010202}
  {\bibfield  {journal} {\bibinfo  {journal} {PRX Quantum}\ }\textbf {\bibinfo
  {volume} {3}},\ \bibinfo {pages} {010202} (\bibinfo {year}
  {2022})}\BibitemShut {NoStop}%
\bibitem [{\citenamefont {Faist}\ \emph {et~al.}(2021)\citenamefont {Faist},
  \citenamefont {Woods}, \citenamefont {Albert}, \citenamefont {Renes},
  \citenamefont {Eisert},\ and\ \citenamefont {Preskill}}]{faist2021time}%
  \BibitemOpen
  \bibfield  {author} {\bibinfo {author} {\bibfnamefont {P.}~\bibnamefont
  {Faist}}, \bibinfo {author} {\bibfnamefont {M.~P.}\ \bibnamefont {Woods}},
  \bibinfo {author} {\bibfnamefont {V.~V.}\ \bibnamefont {Albert}}, \bibinfo
  {author} {\bibfnamefont {J.~M.}\ \bibnamefont {Renes}}, \bibinfo {author}
  {\bibfnamefont {J.}~\bibnamefont {Eisert}},\ and\ \bibinfo {author}
  {\bibfnamefont {J.}~\bibnamefont {Preskill}},\ }\bibfield  {title} {\bibinfo
  {title} {Time-energy uncertainty relation for noisy quantum metrology},\ }in\
  \href@noop {} {\emph {\bibinfo {booktitle} {Quantum Information and
  Measurement}}}\ (\bibinfo {organization} {Optical Society of America},\
  \bibinfo {year} {2021})\ pp.\ \bibinfo {pages} {W2A--3}\BibitemShut {NoStop}%
\bibitem [{\citenamefont {Alipour}\ \emph {et~al.}(2014)\citenamefont
  {Alipour}, \citenamefont {Mehboudi},\ and\ \citenamefont
  {Rezakhani}}]{alipour2014quantum}%
  \BibitemOpen
  \bibfield  {author} {\bibinfo {author} {\bibfnamefont {S.}~\bibnamefont
  {Alipour}}, \bibinfo {author} {\bibfnamefont {M.}~\bibnamefont {Mehboudi}},\
  and\ \bibinfo {author} {\bibfnamefont {A.}~\bibnamefont {Rezakhani}},\
  }\bibfield  {title} {\bibinfo {title} {Quantum metrology in open systems:
  dissipative cram{\'e}r-rao bound},\ }\href@noop {} {\bibfield  {journal}
  {\bibinfo  {journal} {Physical {R}eview {L}etters}\ }\textbf {\bibinfo
  {volume} {112}},\ \bibinfo {pages} {120405} (\bibinfo {year}
  {2014})}\BibitemShut {NoStop}%
\bibitem [{\citenamefont {Naghiloo}\ \emph {et~al.}(2019)\citenamefont
  {Naghiloo}, \citenamefont {Abbasi}, \citenamefont {Joglekar},\ and\
  \citenamefont {Murch}}]{naghiloo2019quantum}%
  \BibitemOpen
  \bibfield  {author} {\bibinfo {author} {\bibfnamefont {M.}~\bibnamefont
  {Naghiloo}}, \bibinfo {author} {\bibfnamefont {M.}~\bibnamefont {Abbasi}},
  \bibinfo {author} {\bibfnamefont {Y.~N.}\ \bibnamefont {Joglekar}},\ and\
  \bibinfo {author} {\bibfnamefont {K.~W.}\ \bibnamefont {Murch}},\ }\bibfield
  {title} {\bibinfo {title} {Quantum state tomography across the exceptional
  point in a single dissipative qubit},\ }\href
  {https://doi.org/10.1038/s41567-019-0652-z} {\bibfield  {journal} {\bibinfo
  {journal} {Nature {P}hysics}\ }\textbf {\bibinfo {volume} {15}},\ \bibinfo
  {pages} {1232} (\bibinfo {year} {2019})}\BibitemShut {NoStop}%
\bibitem [{\citenamefont {Lu}\ \emph {et~al.}(2010)\citenamefont {Lu},
  \citenamefont {Wang},\ and\ \citenamefont {Sun}}]{lu2010quantum}%
  \BibitemOpen
  \bibfield  {author} {\bibinfo {author} {\bibfnamefont {X.-M.}\ \bibnamefont
  {Lu}}, \bibinfo {author} {\bibfnamefont {X.}~\bibnamefont {Wang}},\ and\
  \bibinfo {author} {\bibfnamefont {C.}~\bibnamefont {Sun}},\ }\bibfield
  {title} {\bibinfo {title} {Quantum {F}isher information flow and
  non-markovian processes of open systems},\ }\href@noop {} {\bibfield
  {journal} {\bibinfo  {journal} {Physical Review A}\ }\textbf {\bibinfo
  {volume} {82}},\ \bibinfo {pages} {042103} (\bibinfo {year}
  {2010})}\BibitemShut {NoStop}%
\bibitem [{\citenamefont {Gammelmark}\ and\ \citenamefont
  {M{\o}lmer}(2014)}]{gammelmark2014fisher}%
  \BibitemOpen
  \bibfield  {author} {\bibinfo {author} {\bibfnamefont {S.}~\bibnamefont
  {Gammelmark}}\ and\ \bibinfo {author} {\bibfnamefont {K.}~\bibnamefont
  {M{\o}lmer}},\ }\bibfield  {title} {\bibinfo {title} {Fisher information and
  the quantum cram{\'e}r-rao sensitivity limit of continuous measurements},\
  }\href@noop {} {\bibfield  {journal} {\bibinfo  {journal} {Physical {R}eview
  {L}etters}\ }\textbf {\bibinfo {volume} {112}},\ \bibinfo {pages} {170401}
  (\bibinfo {year} {2014})}\BibitemShut {NoStop}%
\bibitem [{\citenamefont {Altintas}(2016)}]{altintas2016quantum}%
  \BibitemOpen
  \bibfield  {author} {\bibinfo {author} {\bibfnamefont {A.~A.}\ \bibnamefont
  {Altintas}},\ }\bibfield  {title} {\bibinfo {title} {Quantum fisher
  information of an open and noisy system in the steady state},\ }\href@noop {}
  {\bibfield  {journal} {\bibinfo  {journal} {Annals of Physics}\ }\textbf
  {\bibinfo {volume} {367}},\ \bibinfo {pages} {192} (\bibinfo {year}
  {2016})}\BibitemShut {NoStop}%
\bibitem [{\citenamefont {Gorini}\ \emph {et~al.}(1976)\citenamefont {Gorini},
  \citenamefont {Kossakowski},\ and\ \citenamefont
  {Sudarshan}}]{gorini_completely_1976}%
  \BibitemOpen
  \bibfield  {author} {\bibinfo {author} {\bibfnamefont {V.}~\bibnamefont
  {Gorini}}, \bibinfo {author} {\bibfnamefont {A.}~\bibnamefont
  {Kossakowski}},\ and\ \bibinfo {author} {\bibfnamefont {E.~C.~G.}\
  \bibnamefont {Sudarshan}},\ }\bibfield  {title} {\bibinfo {title} {Completely
  positive dynamical semigroups of {N}‐level systems},\ }\href
  {https://doi.org/10.1063/1.522979} {\bibfield  {journal} {\bibinfo  {journal}
  {{Journal of Mathematical Physics}}\ }\textbf {\bibinfo {volume} {17}},\
  \bibinfo {pages} {821} (\bibinfo {year} {1976})},\ \bibinfo {note}
  {publisher: American Institute of Physics}\BibitemShut {NoStop}%
\bibitem [{\citenamefont {Lindblad}(1976)}]{lindblad_generators_1976}%
  \BibitemOpen
  \bibfield  {author} {\bibinfo {author} {\bibfnamefont {G.}~\bibnamefont
  {Lindblad}},\ }\bibfield  {title} {\bibinfo {title} {On the generators of
  quantum dynamical semigroups},\ }\href {https://doi.org/10.1007/BF01608499}
  {\bibfield  {journal} {\bibinfo  {journal} {{Communications in Mathematical
  Physics}}\ }\textbf {\bibinfo {volume} {48}},\ \bibinfo {pages} {119}
  (\bibinfo {year} {1976})}\BibitemShut {NoStop}%
\bibitem [{\citenamefont {Chruściński}\ and\ \citenamefont
  {Pascazio}(2017)}]{chruscinski_brief_2017}%
  \BibitemOpen
  \bibfield  {author} {\bibinfo {author} {\bibfnamefont {D.}~\bibnamefont
  {Chruściński}}\ and\ \bibinfo {author} {\bibfnamefont {S.}~\bibnamefont
  {Pascazio}},\ }\bibfield  {title} {\bibinfo {title} {A {Brief} {History} of
  the {GKLS} {Equation}},\ }\href {https://doi.org/10.1142/S1230161217400017}
  {\bibfield  {journal} {\bibinfo  {journal} {{O}pen {S}ystems \& {I}nformation
  {D}ynamics}\ }\textbf {\bibinfo {volume} {24}},\ \bibinfo {pages} {1740001}
  (\bibinfo {year} {2017})}\BibitemShut {NoStop}%
\bibitem [{\citenamefont {Manzano}(2020)}]{manzano2020short}%
  \BibitemOpen
  \bibfield  {author} {\bibinfo {author} {\bibfnamefont {D.}~\bibnamefont
  {Manzano}},\ }\bibfield  {title} {\bibinfo {title} {A short introduction to
  the {L}indblad master equation},\ }\href@noop {} {\bibfield  {journal}
  {\bibinfo  {journal} {Aip Advances}\ }\textbf {\bibinfo {volume} {10}},\
  \bibinfo {pages} {025106} (\bibinfo {year} {2020})}\BibitemShut {NoStop}%
\bibitem [{\citenamefont {Ashida}\ \emph {et~al.}(2020)\citenamefont {Ashida},
  \citenamefont {Gong},\ and\ \citenamefont {Ueda}}]{ashida2020non}%
  \BibitemOpen
  \bibfield  {author} {\bibinfo {author} {\bibfnamefont {Y.}~\bibnamefont
  {Ashida}}, \bibinfo {author} {\bibfnamefont {Z.}~\bibnamefont {Gong}},\ and\
  \bibinfo {author} {\bibfnamefont {M.}~\bibnamefont {Ueda}},\ }\bibfield
  {title} {\bibinfo {title} {Non-hermitian physics},\ }\href@noop {} {\bibfield
   {journal} {\bibinfo  {journal} {Advances in {P}hysics}\ }\textbf {\bibinfo
  {volume} {69}},\ \bibinfo {pages} {249} (\bibinfo {year} {2020})}\BibitemShut
  {NoStop}%
\bibitem [{\citenamefont {Bender}\ \emph {et~al.}(1999)\citenamefont {Bender},
  \citenamefont {Boettcher},\ and\ \citenamefont {Meisinger}}]{bender1999pt}%
  \BibitemOpen
  \bibfield  {author} {\bibinfo {author} {\bibfnamefont {C.~M.}\ \bibnamefont
  {Bender}}, \bibinfo {author} {\bibfnamefont {S.}~\bibnamefont {Boettcher}},\
  and\ \bibinfo {author} {\bibfnamefont {P.~N.}\ \bibnamefont {Meisinger}},\
  }\bibfield  {title} {\bibinfo {title} {Pt-symmetric quantum mechanics},\
  }\href@noop {} {\bibfield  {journal} {\bibinfo  {journal} {Journal of
  {M}athematical {P}hysics}\ }\textbf {\bibinfo {volume} {40}},\ \bibinfo
  {pages} {2201} (\bibinfo {year} {1999})}\BibitemShut {NoStop}%
\bibitem [{\citenamefont {Lau}\ and\ \citenamefont
  {Clerk}(2018)}]{lau2018fundamental}%
  \BibitemOpen
  \bibfield  {author} {\bibinfo {author} {\bibfnamefont {H.-K.}\ \bibnamefont
  {Lau}}\ and\ \bibinfo {author} {\bibfnamefont {A.~A.}\ \bibnamefont
  {Clerk}},\ }\bibfield  {title} {\bibinfo {title} {Fundamental limits and
  non-reciprocal approaches in non-hermitian quantum sensing},\ }\href@noop {}
  {\bibfield  {journal} {\bibinfo  {journal} {Nature {C}ommunications}\
  }\textbf {\bibinfo {volume} {9}},\ \bibinfo {pages} {1} (\bibinfo {year}
  {2018})}\BibitemShut {NoStop}%
\bibitem [{\citenamefont {Langbein}(2018)}]{Langbein2018}%
  \BibitemOpen
  \bibfield  {author} {\bibinfo {author} {\bibfnamefont {W.}~\bibnamefont
  {Langbein}},\ }\bibfield  {title} {\bibinfo {title} {No exceptional precision
  of exceptional-point sensors},\ }\href@noop {} {\bibfield  {journal}
  {\bibinfo  {journal} {{Physical Review A}}\ }\textbf {\bibinfo {volume}
  {98}},\ \bibinfo {pages} {023805} (\bibinfo {year} {2018})}\BibitemShut
  {NoStop}%
\bibitem [{\citenamefont {Holevo}(2011)}]{holevo2011probabilistic}%
  \BibitemOpen
  \bibfield  {author} {\bibinfo {author} {\bibfnamefont {A.~S.}\ \bibnamefont
  {Holevo}},\ }\href@noop {} {\emph {\bibinfo {title} {Probabilistic and
  statistical aspects of quantum theory}}},\ Vol.~\bibinfo {volume} {1}\
  (\bibinfo  {publisher} {Springer Science \& Business Media},\ \bibinfo {year}
  {2011})\BibitemShut {NoStop}%
\bibitem [{\citenamefont {Helstrom}(1969)}]{helstrom1969quantum}%
  \BibitemOpen
  \bibfield  {author} {\bibinfo {author} {\bibfnamefont {C.~W.}\ \bibnamefont
  {Helstrom}},\ }\bibfield  {title} {\bibinfo {title} {Quantum detection and
  estimation theory},\ }\href@noop {} {\bibfield  {journal} {\bibinfo
  {journal} {Journal of Statistical Physics}\ }\textbf {\bibinfo {volume}
  {1}},\ \bibinfo {pages} {231} (\bibinfo {year} {1969})}\BibitemShut {NoStop}%
\bibitem [{\citenamefont {Liu}\ \emph {et~al.}(2019)\citenamefont {Liu},
  \citenamefont {Yuan}, \citenamefont {Lu},\ and\ \citenamefont
  {Wang}}]{liu2019quantum}%
  \BibitemOpen
  \bibfield  {author} {\bibinfo {author} {\bibfnamefont {J.}~\bibnamefont
  {Liu}}, \bibinfo {author} {\bibfnamefont {H.}~\bibnamefont {Yuan}}, \bibinfo
  {author} {\bibfnamefont {X.-M.}\ \bibnamefont {Lu}},\ and\ \bibinfo {author}
  {\bibfnamefont {X.}~\bibnamefont {Wang}},\ }\bibfield  {title} {\bibinfo
  {title} {Quantum {F}isher information matrix and multiparameter estimation},\
  }\href@noop {} {\bibfield  {journal} {\bibinfo  {journal} {Journal of
  {P}hysics {A}: {M}athematical and {T}heoretical}\ }\textbf {\bibinfo {volume}
  {53}},\ \bibinfo {pages} {023001} (\bibinfo {year} {2019})}\BibitemShut
  {NoStop}%
\bibitem [{\citenamefont {Chen}\ \emph {et~al.}(2021)\citenamefont {Chen},
  \citenamefont {Abbasi}, \citenamefont {Joglekar},\ and\ \citenamefont
  {Murch}}]{chen2021}%
  \BibitemOpen
  \bibfield  {author} {\bibinfo {author} {\bibfnamefont {W.}~\bibnamefont
  {Chen}}, \bibinfo {author} {\bibfnamefont {M.}~\bibnamefont {Abbasi}},
  \bibinfo {author} {\bibfnamefont {Y.~N.}\ \bibnamefont {Joglekar}},\ and\
  \bibinfo {author} {\bibfnamefont {K.~W.}\ \bibnamefont {Murch}},\ }\bibfield
  {title} {\bibinfo {title} {Quantum jumps in the non-hermitian dynamics of a
  superconducting qubit},\ }\href@noop {} {\bibfield  {journal} {\bibinfo
  {journal} {{Physical Review Letters}}\ }\textbf {\bibinfo {volume} {127}},\
  \bibinfo {pages} {140504} (\bibinfo {year} {2021})}\BibitemShut {NoStop}%
\bibitem [{\citenamefont {Zloshchastiev}\ and\ \citenamefont
  {Sergi}(2014)}]{zloshchastiev_comparison_2014}%
  \BibitemOpen
  \bibfield  {author} {\bibinfo {author} {\bibfnamefont {K.~G.}\ \bibnamefont
  {Zloshchastiev}}\ and\ \bibinfo {author} {\bibfnamefont {A.}~\bibnamefont
  {Sergi}},\ }\bibfield  {title} {\bibinfo {title} {Comparison and unification
  of non-{Hermitian} and {Lindblad} approaches with applications to open
  quantum optical systems},\ }\href
  {https://doi.org/10.1080/09500340.2014.930528} {\bibfield  {journal}
  {\bibinfo  {journal} {{Journal of Modern Optics}}\ }\textbf {\bibinfo
  {volume} {61}},\ \bibinfo {pages} {1298} (\bibinfo {year}
  {2014})}\BibitemShut {NoStop}%
\bibitem [{\citenamefont {Cortes}\ \emph {et~al.}(2020)\citenamefont {Cortes},
  \citenamefont {Otten},\ and\ \citenamefont {Gray}}]{cortes2020non}%
  \BibitemOpen
  \bibfield  {author} {\bibinfo {author} {\bibfnamefont {C.~L.}\ \bibnamefont
  {Cortes}}, \bibinfo {author} {\bibfnamefont {M.}~\bibnamefont {Otten}},\ and\
  \bibinfo {author} {\bibfnamefont {S.~K.}\ \bibnamefont {Gray}},\ }\bibfield
  {title} {\bibinfo {title} {Non-hermitian approach for quantum plasmonics},\
  }\href@noop {} {\bibfield  {journal} {\bibinfo  {journal} {{The Journal of
  Chemical Physics}}\ }\textbf {\bibinfo {volume} {152}},\ \bibinfo {pages}
  {084105} (\bibinfo {year} {2020})}\BibitemShut {NoStop}%
\bibitem [{\citenamefont {Boixo}\ \emph {et~al.}(2007)\citenamefont {Boixo},
  \citenamefont {Flammia}, \citenamefont {Caves},\ and\ \citenamefont
  {Geremia}}]{boixo_generalized_2007}%
  \BibitemOpen
  \bibfield  {author} {\bibinfo {author} {\bibfnamefont {S.}~\bibnamefont
  {Boixo}}, \bibinfo {author} {\bibfnamefont {S.~T.}\ \bibnamefont {Flammia}},
  \bibinfo {author} {\bibfnamefont {C.~M.}\ \bibnamefont {Caves}},\ and\
  \bibinfo {author} {\bibfnamefont {J.~M.}\ \bibnamefont {Geremia}},\
  }\bibfield  {title} {\bibinfo {title} {Generalized {Limits} for
  {Single}-{Parameter} {Quantum} {Estimation}},\ }\href
  {https://doi.org/10.1103/PhysRevLett.98.090401} {\bibfield  {journal}
  {\bibinfo  {journal} {Phys. Rev. Lett.}\ }\textbf {\bibinfo {volume} {98}},\
  \bibinfo {pages} {090401} (\bibinfo {year} {2007})}\BibitemShut {NoStop}%
\bibitem [{\citenamefont {Giovannetti}\ \emph
  {et~al.}(2004{\natexlab{b}})\citenamefont {Giovannetti}, \citenamefont
  {Lloyd},\ and\ \citenamefont {Maccone}}]{lloyd2004}%
  \BibitemOpen
  \bibfield  {author} {\bibinfo {author} {\bibfnamefont {V.}~\bibnamefont
  {Giovannetti}}, \bibinfo {author} {\bibfnamefont {S.}~\bibnamefont {Lloyd}},\
  and\ \bibinfo {author} {\bibfnamefont {L.}~\bibnamefont {Maccone}},\
  }\bibfield  {title} {\bibinfo {title} {Quantum-enhanced measurements:
  {B}eating the standard quantum limit},\ }\href@noop {} {\bibfield  {journal}
  {\bibinfo  {journal} {Science}\ }\textbf {\bibinfo {volume} {306}},\ \bibinfo
  {pages} {1330} (\bibinfo {year} {2004}{\natexlab{b}})}\BibitemShut {NoStop}%
\bibitem [{\citenamefont {Campos-Gonzalez-Angulo}\ and\ \citenamefont
  {Yuen-Zhou}(2022)}]{Campos}%
  \BibitemOpen
  \bibfield  {author} {\bibinfo {author} {\bibfnamefont {J.~A.}\ \bibnamefont
  {Campos-Gonzalez-Angulo}}\ and\ \bibinfo {author} {\bibfnamefont
  {J.}~\bibnamefont {Yuen-Zhou}},\ }\bibfield  {title} {\bibinfo {title}
  {Generalization of the {T}avis-{C}ummings model for multi-level anharmonic
  systems: Insights on the second excitation manifold},\ }\href@noop {}
  {\bibfield  {journal} {\bibinfo  {journal} {J. Chem. Phys.}\ }\textbf
  {\bibinfo {volume} {156}},\ \bibinfo {pages} {194308} (\bibinfo {year}
  {2022})}\BibitemShut {NoStop}%
\bibitem [{\citenamefont {Tavis}\ and\ \citenamefont {Cummings}(1968)}]{Tavis}%
  \BibitemOpen
  \bibfield  {author} {\bibinfo {author} {\bibfnamefont {M.}~\bibnamefont
  {Tavis}}\ and\ \bibinfo {author} {\bibfnamefont {F.~W.}\ \bibnamefont
  {Cummings}},\ }\bibfield  {title} {\bibinfo {title} {Exact solution for an
  {N}-molecule-radiation-field {H}amiltonian},\ }\href@noop {} {\bibfield
  {journal} {\bibinfo  {journal} {Physical Review}\ }\textbf {\bibinfo {volume}
  {170}},\ \bibinfo {pages} {379} (\bibinfo {year} {1968})}\BibitemShut
  {NoStop}%
\bibitem [{\citenamefont {Garraway}(2011)}]{Garraway}%
  \BibitemOpen
  \bibfield  {author} {\bibinfo {author} {\bibfnamefont {B.~M.}\ \bibnamefont
  {Garraway}},\ }\bibfield  {title} {\bibinfo {title} {The {D}icke model in
  quantum optics: {D}icke model revisited},\ }\href@noop {} {\bibfield
  {journal} {\bibinfo  {journal} {Phil. Trans. R. Soc. A}\ }\textbf {\bibinfo
  {volume} {369}},\ \bibinfo {pages} {1137} (\bibinfo {year}
  {2011})}\BibitemShut {NoStop}%
\bibitem [{\citenamefont {Sidhu}\ and\ \citenamefont {Kok}(2020)}]{kok2020}%
  \BibitemOpen
  \bibfield  {author} {\bibinfo {author} {\bibfnamefont {J.}~\bibnamefont
  {Sidhu}}\ and\ \bibinfo {author} {\bibfnamefont {P.}~\bibnamefont {Kok}},\
  }\bibfield  {title} {\bibinfo {title} {Geometric perspective on quantum
  parameter estimation},\ }\href@noop {} {\bibfield  {journal} {\bibinfo
  {journal} {{AVS Quantum Science}}\ }\textbf {\bibinfo {volume} {2}},\
  \bibinfo {pages} {014701} (\bibinfo {year} {2020})}\BibitemShut {NoStop}%
\end{thebibliography}%

\end{document}